\documentclass{article}

\usepackage{iclr2027_conference,times}

\usepackage[utf8]{inputenc}
\usepackage[T1]{fontenc}
\usepackage{microtype}

\usepackage{amsmath,amssymb,amsfonts}
\usepackage{pifont}
\usepackage{tablefootnote}

\usepackage{graphicx}
\usepackage[table]{xcolor}
\usepackage{booktabs}
\usepackage{tabularx}
\usepackage{array}
\usepackage{multirow}
\usepackage{longtable}
\usepackage{arydshln}

\usepackage{algorithm}
\usepackage{enumitem}

\usepackage{url}
\usepackage{hyperref}

\usepackage{booktabs}
\usepackage{threeparttable}
\usepackage{tabularx}
\usepackage{makecell}

\usepackage[most]{tcolorbox}

\definecolor{PromptBg}{HTML}{F8FAFC}
\definecolor{PromptFrame}{HTML}{D7DFE8}
\definecolor{PromptTitleBg}{HTML}{EAF0F7}
\definecolor{PromptTitle}{HTML}{243B53}

\newtcolorbox{agentprompt}[1]{
    enhanced,
    breakable,
    colback=PromptBg,
    colframe=PromptFrame,
    colbacktitle=PromptTitleBg,
    coltitle=PromptTitle,
    title={#1},
    fonttitle=\bfseries\sffamily\small,
    fontupper=\small\ttfamily\raggedright,
    boxrule=0.5pt,
    arc=1.5mm,
    left=10pt,
    right=10pt,
    top=9pt,
    bottom=9pt,
    before skip=8pt,
    after skip=10pt
}

\title{
Thought Blocks: Non-Autoregressive Latent Communication for Multi-Agent Systems via Trajectory-to-State Compilation
}

\author{%
\makebox[\textwidth][c]{%
\shortstack[c]{%
\\[5pt]
\textbf{Zhenhan Guan}$^{1,2}$ \quad
\textbf{Xiaolong Liu}$^{1,2}$ \quad
\textbf{Hengwei Ye}$^{1}$ \quad
\textbf{Zheng Tian}$^{1,*}$
\\[10pt]
{\normalfont\small
$^{1}$ ShanghaiTech University, Shanghai, China}
\\
{\normalfont\small
$^{2}$ AI Lab, The Yangtze River Delta, Changzhou, China}
\\[8pt]
{\normalfont\footnotesize
\texttt{\{guanzhh2025,liuxl2025,yehw2024\}@shanghaitech.edu.cn}}
\\[3pt]
{\normalfont\footnotesize
$^{*}$ Corresponding author: \texttt{tianzheng@shanghaitech.edu.cn}}
}%
}%
}

\iclrfinalcopy

\begin{document}
\newcommand{\cmark}{\ding{51}}
\newcommand{\xmark}{\ding{55}}
\maketitle

\begin{abstract}

LLM-based multi-agent systems typically communicate through text,
incurring repeated generation and re-encoding costs.
Although latent communication removes explicit verbalization,
constructing latent messages may still require autoregressive
latent trajectories.
We therefore distinguish \emph{message representation} from
\emph{message construction} and ask whether agents can directly
construct useful continuation states without executing the
trajectories that normally produce them.
We formulate this problem as \emph{trajectory-to-state compilation}
and introduce \emph{Thought Blocks}, fixed-size blocks of latent
key--value (KV) continuation memory constructed in one shot. Thought Blocks achieve a $2.5\times$ average latency speedup over autoregressive latent communication.
Compared with text-based multi-agent communication,
Thought Blocks reduce text tokens by $45.7\%$ and autoregressive
steps by $84.8\%$, achieving an $8.6\times$ average latency
speedup.
These results demonstrate that trajectory-to-state compilation enables efficient inter-agent latent communication without sequential latent rollouts at inference.
\end{abstract}

\section{Introduction}
\label{sec:introduction}

LLM-based multi-agent systems have emerged as a promising paradigm for solving complex tasks through decomposition, specialization, and iterative refinement
\citep{li2023camel,wu2023autogen,hong2023metagpt,qian2024chatdev,du2023improving,liu2023dylan}.
As these systems are applied to increasingly long-horizon and computation-intensive workflows, communication between agents becomes an important systems concern.
For example, OpenAI recently reported using a system of coordinating agents in its work on the Navier--Stokes Millennium Prize Problem, involving approximately 2.7 million inter-agent messages and 130 billion output tokens
\citep{openai2026navierstokes}.
At such scales, the cost of repeatedly constructing and exchanging intermediate information can become a substantial part of multi-agent computation.

Most existing multi-agent systems communicate through natural language
\citep{qian2024chatdev,yan2025beyond}.
Recent latent-communication methods challenge this interface by showing that intermediate information can instead be carried through model-internal representations
\citep{zou2025latent,du2025interlat,ramesh2025communicating}.
More importantly, recent mechanistic evidence from the Jacobian Lens identifies a small internal ``J-space'' in which intermediate results can be retained and reused by subsequent computation
\citep{gurnee2026verbalizable}.
This suggests a separation between the trajectory that produces an intermediate result and the state through which it becomes available to later computation.
For inter-agent communication, this motivates a natural hypothesis:
a receiver may need only a functionally sufficient continuation state, rather than the sender's full latent computation trajectory.

This possibility highlights a key distinction: message representation and message construction are separate design choices, and a message can be latent while its construction remains autoregressive.
For example, some latent multi-agent protocols eliminate textual messages but still generate a sequence of latent thoughts before transferring the resulting latent state to the next agent
\citep{zou2025latent,li2026lesslatent}.
The communicated object is latent, yet constructing it still requires a sequential latent rollout.
We therefore ask:
\emph{must a useful inter-agent latent state be obtained by executing the trajectory that normally produces it, or can it be constructed directly from the sender's current state?}
We formulate this as \emph{non-autoregressive latent communication (NALC)}:
constructing inter-agent latent messages without a sequential latent rollout.
We call the underlying principle \emph{trajectory-to-state compilation}:
directly constructing the continuation state that the rollout would otherwise produce.

We instantiate this idea with \emph{Thought Blocks}, fixed-size blocks of latent KV continuation memory constructed in one shot.
Our method represents the sender using a factorized continuation state consisting of a compact \emph{continuation seed}, which indicates how computation should continue, and a \emph{context bank}, which preserves contextual information that subsequent computation may need to access.
A Thought Block writer maps this state directly to an outgoing KV block without autoregressive intermediate-message generation.
To train the writer, we use an autoregressive latent teacher to identify useful continuation states and distill these state-level targets into the one-shot writer.
The teacher trajectory is therefore used only for supervision during training and is removed entirely from inter-agent message construction at inference time.

Experiments across reasoning and question-answering benchmarks and multiple LLM backbones show that direct continuation-state construction provides a favorable accuracy--efficiency trade-off.
Compared with text-based multi-agent communication, Thought Blocks reduce text tokens by 45.7\%, autoregressive forward steps by 84.8\%, and end-to-end latency by 8.6$\times$ on average, with an average accuracy difference of only $-0.15$ points.
More importantly, compared with autoregressive latent communication, where both methods already communicate in latent space, Thought Blocks further reduce autoregressive forward steps by 35.7\%, with an average accuracy difference of $-0.72$ points.
These results provide empirical evidence that useful inter-agent continuation memory need not be constructed by replaying an autoregressive latent trajectory. Our contributions are threefold:
\begin{itemize}[leftmargin=1.2em, itemsep=0.25em, topsep=0.25em]

\item \textbf{We formulate non-autoregressive latent communication as a message-construction problem.}
We distinguish the representation of an inter-agent message from the process used to construct it, and introduce \emph{trajectory-to-state compilation} as directly constructing a useful continuation state without executing the trajectory that would normally produce it.

\item \textbf{We introduce Thought Blocks for one-shot latent communication.}
Thought Blocks instantiate trajectory-to-state compilation by directly constructing fixed-size KV continuation memory from a factorized sender state and are trained via state-level distillation from autoregressive latent trajectories.

\item \textbf{We show that autoregressive latent message construction can be substantially reduced while largely preserving downstream utility.}
Across multiple backbones and diverse reasoning and question-answering tasks, Thought Blocks substantially reduce sequential communication computation, latency, and KV-memory load while maintaining competitive task performance.

\end{itemize}

\section{Related Work}

\label{sec:related_work}

\subsection{Efficient Multi-Agent Communication}

LLM-based multi-agent frameworks typically coordinate specialized roles
through natural-language interaction~\citep{li2023camel,
wu2023autogen,hong2023metagpt,qian2024chatdev}.
Work on communication efficiency reduces this textual overhead through
learned communication policies~\citep{chen2024optima} or by pruning
redundant agents and links~\citep{wang2025agentdropout}, while retaining
natural language as the communication medium.

A complementary line of work communicates through model-internal
representations. Activation Communication (AC) exchanges intermediate
activations~\citep{ramesh2025communicating}; InterLat communicates
hidden states and studies latent compression~\citep{du2025interlat};
and Thought Communication extracts latent factors for
collaboration~\citep{zheng2025thought}.
At the KV-cache level, KVCOMM reuses and adapts KV states across
contexts~\citep{ye2025kvcomm}, KVComm selectively shares informative
KV pairs~\citep{shi2026kvcomm}, C2C projects and fuses source and
receiver caches~\citep{fu2025cache}, and LCF learns compact cache
channels~\citep{rossi2026latentcacheflow}.
StateBridge aligns sender hidden states to the receiver space without
training~\citep{peng2026statebridge}, while recent work studies
heterogeneous KV communication through dense cross-model cache
alignment~\citep{chen2026see} and dual-cache transformation
~\citep{liu2026dualcache}.
Relayed KV memories can also be compressed while preserving
task-relevant information~\citep{li2026lesslatent}.

These methods improve how model-internal states are selected, aligned,
transferred, or compressed. In particular, cache-transfer methods
primarily operate on sender states that have already been materialized.
Our setting instead targets a continuation state that is not yet
available at the communication boundary and would ordinarily be reached
only after additional latent computation.
We ask whether this post-trajectory state can be constructed directly
from the pre-trajectory sender state, without executing that trajectory.
Accordingly, we use \emph{non-autoregressive latent communication} to
refer specifically to eliminating trajectory-based message construction,
rather than broadly to any latent protocol that avoids token-by-token
text decoding.

\subsection{Latent Trajectories and Direct State Construction}

\begin{table*}[t]
\centering
\caption{Comparison of latent reasoning and communication mechanisms.}
\label{tab:mechanistic_comparison}
\scriptsize
\setlength{\tabcolsep}{4.2pt}
\renewcommand{\arraystretch}{1.08}

\begin{tabular}{lccccc}
\toprule
Method
& \shortstack{Inter-agent\\communication}
& \shortstack{KV\\message}
& \shortstack{Sequential\\latent rollout}
& \shortstack{Iterative latent\\refinement}
& \shortstack{Direct post-trajectory\\state} \\

\midrule
\multicolumn{6}{l}{\textit{Single-agent latent reasoning}} \\
Coconut \cite{hao2024training}
    & \xmark & \xmark & \cmark & \xmark & \xmark \\
PCCoT \cite{wu2025parallel}
    & \xmark & \xmark & \xmark & \cmark & \xmark \\
CoT2 \cite{gozeten2026cot2}
    & \xmark & \xmark & \cmark & \xmark & \xmark \\
LaDiR \cite{kang2025ladir}
    & \xmark & \xmark & \xmark & \cmark & \xmark \\

\midrule
\multicolumn{6}{l}{\textit{Trajectory-produced latent communication}} \\
InterLat-Comp. \cite{du2025interlat}
    & \cmark & \xmark & \cmark & \xmark & \xmark \\
LessLatent \cite{li2026lesslatent}
    & \cmark & \cmark & \cmark & \xmark & \xmark \\
LatentMAS \cite{zou2025latent}
    & \cmark & \cmark & \cmark & \xmark & \xmark \\

\midrule
\multicolumn{6}{l}{\textit{Latent-state transfer / transformation}} \\
AC \cite{ramesh2025communicating}
    & \cmark & \xmark & \xmark & \xmark & \xmark \\
InterLat \cite{du2025interlat}
    & \cmark & \xmark & \xmark & \xmark & \xmark \\
Thought Communication \cite{zheng2025thought}
    & \cmark & \xmark & \xmark & \xmark & \xmark \\
KVCOMM (Ye et al.) \cite{ye2025kvcomm}
    & \cmark & \cmark & \xmark & \xmark & \xmark \\
KVComm (Shi et al.) \cite{shi2026kvcomm}
    & \cmark & \cmark & \xmark & \xmark & \xmark \\
C2C \cite{fu2025cache}
    & \cmark & \cmark & \xmark & \xmark & \xmark \\
LCF \cite{rossi2026latentcacheflow}
    & \cmark & \cmark & \xmark & \xmark & \xmark \\
StateBridge \cite{peng2026statebridge}
    & \cmark & \xmark & \xmark & \xmark & \xmark \\
XKV \cite{liu2026dualcache}
    & \cmark & \cmark & \xmark & \xmark & \xmark \\

\midrule
\textbf{Thought Blocks (Ours)}
    & \cmark & \cmark & \xmark & \xmark & \cmark \\
\bottomrule
\end{tabular}
\end{table*}

Latent representations do not by themselves eliminate sequential
computation.
Coconut recurrently feeds hidden states back as continuous
inputs~\citep{hao2024training}; LatentMAS autoregressively generates
latent thoughts before relaying the resulting working
memory~\citep{zou2025latent}; and InterLat's compressed variant likewise
uses recurrent latent computation~\citep{du2025interlat}.
Thus, latent messages may still require a sequence of intermediate
states to construct.

Other work reduces or reorganizes this sequential structure.
PCCoT uses Jacobi iteration to update multiple continuous thought
positions in parallel~\citep{wu2025parallel}; CoT2 tracks multiple
reasoning alternatives in parallel within each decoding
step~\citep{gozeten2026cot2}; and LaDiR constructs latent thought blocks
through iterative diffusion-based denoising~\citep{kang2025ladir}.
These methods parallelize or iteratively refine latent computation, but
inference still instantiates intermediate latent states before the
resulting reasoning state becomes available.

We instead ask whether such a trajectory needs to be executed at all
when its purpose is to produce a state for another agent.
This motivates \emph{trajectory-to-state compilation}: during training,
an autoregressive teacher identifies a useful post-trajectory
continuation state, while the student learns to map the pre-trajectory
sender state directly to that target.
At inference, the teacher rollout is omitted and the outgoing KV
continuation memory is constructed in one shot.
This separates our work from both transfer or transformation of
already-materialized states and methods that accelerate latent
trajectory construction. Table~\ref{tab:mechanistic_comparison} summarizes this distinction:
Thought Blocks perform inter-agent KV communication by directly
constructing a post-trajectory continuation state, without sequential
latent rollout or iterative refinement at inference.

\section{Method}
\label{sec:method}

\subsection{Problem Setup}
\label{sec:problem_setup}

We consider sequential collaboration among $N$ language-model agents
$\{A_i\}_{i=1}^{N}$.
Given a task $q$, each agent $A_i$ receives a local textual input $x_i$
and, except for the first agent, an inter-agent message
$\mathcal{M}_{i-1}$ from its predecessor, with
$\mathcal{M}_0=\varnothing$.
The local input includes the agent's role instruction and may additionally
contain the original task, retrieved evidence, or other observations.

Let $\mathcal{S}_i$ denote the sender-side state available after agent
$A_i$ has processed its local input and incoming message.
A generic communication step can therefore be written as
\begin{equation}
\mathcal{S}_i
=
A_i(x_i,\mathcal{M}_{i-1}),
\qquad
\mathcal{M}_i
=
\Gamma_i(\mathcal{S}_i),
\quad i<N,
\label{eq:generic_mas}
\end{equation}
where $\Gamma_i$ denotes the mechanism that constructs the outgoing
message from the sender state.
The final agent consumes its local input and the last incoming message to
produce the task output $\hat{y}
=
A_N(x_N,\mathcal{M}_{N-1})$. Different communication methods instantiate $\mathcal{M}_i$ and
$\Gamma_i$ differently.
Text-based systems verbalize intermediate computation into natural-language
tokens, whereas latent-communication systems expose model-internal states
directly.
In this work, we represent the communicated object as latent KV
continuation memory and focus on a separate question:
\emph{how should this latent message be constructed?}

In trajectory-based latent communication, the outgoing state is obtained
by first executing a sequential latent computation,
\begin{equation}
\tau_i
=
\mathcal{T}_{\theta^\star}(\mathcal{S}_i),
\qquad
\mathcal{M}_i^\star
=
\Psi_{\theta^\star}(\tau_i),
\label{eq:trajectory_construction}
\end{equation}
where
$\tau_i=(u_{i,1},\ldots,u_{i,T_{\mathrm{lat}}})$
is a latent trajectory,
$\mathcal{T}_{\theta^\star}$ denotes the sequential latent-computation
process, and $\Psi_{\theta^\star}$ extracts the continuation memory
communicated to the next agent.
Although $\mathcal{M}_i^\star$ is latent, constructing it still requires
executing the trajectory $\tau_i$. Our goal is to replace this trajectory-dependent construction with a
direct parameterized writer,
\begin{equation}
\mathcal{M}_i
=
\mathcal{W}_{\phi;\theta^\star}(\mathcal{S}_i),
\label{eq:direct_construction}
\end{equation}
where $\phi$ denotes trainable writer parameters and $\theta^\star$
denotes the frozen language-model backbone.
We refer to this replacement of latent-trajectory execution with direct
continuation-state construction as
\emph{trajectory-to-state compilation}.

The remainder of this section first specifies the continuation state
$\mathcal{S}_i$, then describes how Thought Blocks directly construct an
outgoing latent message from this state, and finally explains how this
direct mapping is learned from autoregressive trajectories.
At inference time, Eq.~\ref{eq:direct_construction} is applied by each
intermediate agent, while the final textual answer is generated through the
standard autoregressive language-model interface.

\subsection{Factorized Continuation State}
\label{sec:factorized_state}

Direct continuation construction requires preserving two distinct forms of
information available after prefill: a boundary representation that
conditions subsequent computation, and addressable prefix memory from which
future states can retrieve context. We therefore factorize the sender state as
\begin{equation}
\mathcal{S}_i
=
(R_i,C_i),
\label{eq:factorized_state}
\end{equation}
where $R_i$ is the \emph{continuation seed} and $C_i$ is the
\emph{context bank}.

\paragraph{Continuation seed.}
Let $F$ denote the final prefill position after agent $A_i$ processes its
local input $x_i$ together with the incoming latent message
$\mathcal{M}_{i-1}$. We form the continuation seed by stacking the hidden
states at this boundary across all $L$ backbone layers:
\begin{equation}
R_i
=
\left[
h_{i,F}^{1};
h_{i,F}^{2};
\ldots;
h_{i,F}^{L}
\right]
\in
\mathbb{R}^{L\times d},
\label{eq:continuation_seed}
\end{equation}
where $d$ is the hidden dimension.
Unlike a final-layer-only readout, $R_i$ exposes the writer to the full
layer-wise representation at the continuation boundary, providing a compact
summary from which the outgoing continuation can be constructed.

\paragraph{Context bank.}
The boundary representation alone does not provide the addressable prefix
memory used by Transformer attention. We therefore retain a compressed KV
context bank,
\begin{equation}
C_i
=
\left\{
\left(K_{i,n}^{\ell},V_{i,n}^{\ell}\right)
\right\}_{\substack{
\ell=1,\ldots,L\\
n=1,\ldots,N_i^{C}
}},
\label{eq:context_bank}
\end{equation}
where $N_i^{C}$ is the number of retained context positions.
Thus, $R_i$ summarizes the computation at the continuation boundary, while
$C_i$ preserves contextual information in an explicitly retrievable form.
This factorization allows the writer to condition continuation construction
on $R_i$ without requiring the seed itself to encode the entire prefix. In our implementation, $C_i$ is instantiated using Activation
Beacon~\citep{zhang2024activation}. Details of beacon training, compression,
and target extraction are provided in
Appendix~\ref{app:thought_block_training}.

\subsection{One-Shot Thought Block Writer}
\label{sec:thought_block_writer}

Autoregressive continuation uses $R_i$ and $C_i$ through a sequential
process.
Starting from the continuation boundary, each newly generated latent state
becomes part of the computation used to construct subsequent states, while
the prefix memory remains accessible throughout the rollout.
The resulting trajectory therefore performs two functions jointly:
it \emph{expands} the current computational state into future positions and
\emph{contextualizes} those positions using the available prefix memory.

Once the autoregressive rollout is removed, these two functions must be
implemented differently.
Thought Blocks separate them into
\emph{parallel continuation expansion} and
\emph{memory-backed rendering}.

\paragraph{Parallel continuation expansion.}

Autoregressive decoding obtains multiple future positions through temporal
recurrence.
Without this recurrence, a single continuation seed must instead be
expanded directly into multiple latent positions.
We introduce $B$ learnable slots
\begin{equation}
O
=
[o_1,o_2,\ldots,o_B]
\in
\mathbb{R}^{B\times d_s},
\label{eq:thought_slots}
\end{equation}
where each slot indexes one position in the outgoing Thought Block. The slots act as position-specific queries over the layer-wise continuation
seed $R_i$. We construct $B$ sender-conditioned block embeddings in parallel
through a lightweight cross-attention module,
\begin{equation}
E_i
=
g_\phi(O,R_i)
\in
\mathbb{R}^{B\times d},
\label{eq:block_expansion}
\end{equation}
where $O$ provides the queries and $R_i$ provides the keys and values.

Equation~\ref{eq:block_expansion} replaces
\emph{temporal state expansion} with
\emph{parallel positional expansion}.
Rather than producing future positions through recurrent state transitions,
the $B$ slots query the same continuation boundary simultaneously.
Because $R_i$ contains boundary representations from all backbone layers,
different slots can extract different combinations of layer-wise information,
yielding distinct sender-conditioned block embeddings without recurrence
across block positions.

Importantly, this expansion is intentionally driven by $R_i$ rather than
requiring $E_i$ to encode the entire preceding context.
The role of $E_i$ is to specify \emph{where the computation should
continue}, not to reproduce everything the model may subsequently need to
remember.

\paragraph{Memory-backed rendering.}
The expanded block embeddings $E_i$ specify $B$ candidate continuation
positions, but they do not yet constitute model-native continuation memory.
We therefore render them through the frozen backbone while retaining access
to the context bank $C_i$. Specifically, we treat $E_i$ as the input embeddings of $B$ virtual
continuation positions and perform a single causal Transformer forward pass.
At each layer, the retained keys and values in $C_i$ are supplied as prefix
KV memory, allowing the block positions to retrieve information from the
compressed prefix through the backbone's native self-attention.
The complete writer is therefore
\begin{equation}
\mathcal{M}_i
=
\mathcal{W}_{\phi;\theta^\star}(R_i,C_i)
=
\mathrm{Render}_{\theta^\star}(E_i;C_i),
\qquad
E_i=g_\phi(O,R_i),
\label{eq:thought_block_writer}
\end{equation}
where $\mathrm{Render}_{\theta^\star}$ denotes this single frozen-backbone
forward pass. During rendering, the Transformer materializes a key and value at every
block position and every layer. We collect these states as the outgoing
latent message,
\begin{equation}
\mathcal{M}_i
=
\left\{
\left(
\widehat{K}_{i,j}^{\ell},
\widehat{V}_{i,j}^{\ell}
\right)
\right\}_{\substack{
\ell=1,\ldots,L\\
j=1,\ldots,B
}}.
\label{eq:thought_block_memory}
\end{equation}
Although the $B$ continuation positions interact under the backbone's
standard causal attention mask, they are processed within a single forward
pass rather than generated through $B$ autoregressive decode steps.

\subsection{Trajectory-to-State Distillation}
\label{sec:trajectory_distillation}

Equation~\ref{eq:thought_block_writer} specifies how to construct
continuation memory directly.
We now describe how the writer learns which continuation state to
construct.

\paragraph{Teacher state construction.}

Starting from the same factorized continuation state $(R_i,C_i)$, the
frozen backbone performs autoregressive latent decoding following the
continuous latent-feedback interface used in autoregressive latent
communication~\citep{zou2025latent}:
\begin{equation}
\tau_i
=
[u_{i,1},\ldots,u_{i,T_{\mathrm{lat}}}]
=
\mathrm{ARDecode}_{\theta^\star}(R_i;C_i).
\label{eq:teacher_rollout}
\end{equation}

The trajectory contains a sequence of intermediate states, whereas the
student communicates a fixed-size block.
We therefore apply a state-extraction operator $\Pi$ to obtain a
fixed-size teacher target:
\begin{equation}
\mathcal{M}_i^\star
=
\Pi(\tau_i).
\label{eq:teacher_state}
\end{equation}

Importantly, the student is not trained to reproduce the trajectory
$\tau_i$ itself.
The trajectory serves only as the computational path through which the
teacher reaches $\mathcal{M}_i^\star$.
The target of distillation is the resulting continuation state.

\paragraph{Direct continuation-state reconstruction.}

The student receives only the pre-trajectory state $(R_i,C_i)$ and directly constructs $\widehat{\mathcal{M}}_i = \mathcal{W}_{\phi;\theta^\star}(R_i,C_i)$. Let
$\widehat{K}_{i,j}^{\ell},
\widehat{V}_{i,j}^{\ell}$
denote the student KV states at block position $j$ and layer $\ell$, and
let
$K_{i,j}^{\ell,\star},
V_{i,j}^{\ell,\star}$
denote the corresponding teacher targets.
The primary distillation objective is
\begin{equation}
\mathcal{L}_{\mathrm{KV}}
=
\frac{1}{BL}
\sum_{\ell=1}^{L}
\sum_{j=1}^{B}
\left(
\left\|
\widehat{K}_{i,j}^{\ell}
-
K_{i,j}^{\ell,\star}
\right\|_F^2
+
\left\|
\widehat{V}_{i,j}^{\ell}
-
V_{i,j}^{\ell,\star}
\right\|_F^2
\right).
\label{eq:kv_loss}
\end{equation}

This objective directly matches the representation consumed by the
downstream agent.
The teacher constructs the target continuation
states through sequential latent decoding, whereas the student learns to
construct it directly from the continuation boundary state. We additionally use an auxiliary hidden-state alignment objective during
training.
Because this auxiliary state is neither communicated nor required at
inference, its definition is deferred to
Appendix~\ref{app:thought_block_training}.
The full objective is
\begin{equation}
\mathcal{L}_{\mathrm{TSD}}
=
\lambda_{\mathrm{KV}}\mathcal{L}_{\mathrm{KV}}
+
\lambda_{\mathrm{aux}}\mathcal{L}_{\mathrm{aux}}.
\label{eq:tsd_loss}
\end{equation}

Only the writer parameters $\phi$ are optimized during this stage; the
language-model backbone and contextual-memory mechanism remain fixed.
Trajectory-to-state distillation therefore induces an asymmetric
teacher--student computation:
the teacher executes a sequential trajectory to reach a useful continuation state, while the student learns the direct mapping $(R_i,C_i)
\longrightarrow
\widehat{\mathcal{M}}_i.$
At inference time, the teacher trajectory and auxiliary training signal are
removed entirely.

\section{Experiments}





\subsection{Experimental Setup}
\label{sec:experimental_setup}

\paragraph{Tasks and datasets.}
We evaluate on eight datasets spanning three task categories:
(1) \textbf{Science, Commonsense, and Mathematical Reasoning}, including ARC~\citep{clark2018think}, OpenBookQA~\citep{mihaylov2018can}, and GSM8K~\citep{cobbe2021training};
(2) \textbf{Knowledge-Intensive Question Answering}, including MMLU~\citep{hendrycks2021measuring}, GPQA~\citep{rein2024gpqa}, and MedQA~\citep{jin2021disease};
and (3) \textbf{Reading Comprehension}, including BoolQ~\citep{clark2019boolq} and QuAIL~\citep{rogers2020getting}.
We evaluate ARC on both the Challenge and Easy splits and MedQA on its US, Taiwan, and Mainland variants, yielding 11 benchmark configurations in total.

\paragraph{Baselines.}
We compare against three baselines.
(1) \textbf{Single-agent}: a single model using the same backbone and a matched total decoding budget.
(2) \textbf{Text-based multi-agent (Text-MAS)}: agents follow the same role decomposition as our method but exchange natural-language messages.
(3) \textbf{Autoregressive latent communication (Latent-MAS)}: intermediate agents communicate through autoregressively generated latent thoughts~\citep{zou2025latent}.
All multi-agent methods use the same backbone, role decomposition, and number of agents, so they differ primarily in their inter-agent communication mechanism.

\paragraph{Metrics.}

We evaluate task utility and inference efficiency, distinguishing
\emph{end-to-end} from \emph{communication-stage} cost.
Task performance is measured by \textbf{accuracy}, using the official
evaluation protocol where available and otherwise deterministic exact
matching against official gold labels. For end-to-end efficiency, we report
\textbf{text tokens}, \textbf{AR Steps}, and \textbf{latency}.
Text tokens count the tokenizer-visible input and generated output tokens processed across all agents; AR Steps count autoregressive decoding steps across intermediate
communication and final-answer generation; and latency measures wall-clock
time for the full pipeline, including prefill, communication, and final decoding.

To capture sequential communication cost not reflected by token counts, we
also report \textbf{communication forward passes (Passes)}, defined as
the number of intermediate-agent forward passes used to construct outgoing
messages. Text-MAS and Latent-MAS require sequential passes, whereas Thought
Blocks use one non-autoregressive rendering pass per intermediate agent and
therefore incur zero autoregressive communication steps. We additionally
report \textbf{KV-cache load}, the amount of KV state loaded during decoding.
All latency measurements use the same hardware and configuration and
implementation details are provided in Appendix~\ref{app:implementation_details}.

\subsection{Main Results}

Table~\ref{tab:main_results_category_average} compares Thought Blocks
with text-based and autoregressive latent multi-agent communication
across four LLM backbones\footnote{Full results across all benchmarks
are provided in Appendix Table~\ref{tab:full_main_results}.}.
Overall, Thought Blocks retain comparable task performance while
substantially reducing sequential communication and end-to-end
decoding cost. Relative to Latent-MAS, the more direct comparison for isolating message
construction in latent space, Thought Blocks reduce communication passes
from $12$ to $3$ under our four-agent setup ($75.0\%$), reduce end-to-end
autoregressive steps by $35.7\%$ on average, and achieve a $2.5\times$
average latency speedup, with an average accuracy difference of
$-0.72$ points. Relative to Text-MAS, Thought Blocks reduce text tokens
by $45.7\%$, communication passes by $99.4\%$, and end-to-end
autoregressive steps by $84.8\%$, while achieving an $8.6\times$
average latency speedup with an average accuracy difference of only
$-0.15$ points. Compared with the decoding-budget-matched Single Text
baseline, Thought Blocks achieve higher accuracy in all $12$
backbone--category settings, with a $+4.25$-point average improvement.

These comparisons separate two sources of communication overhead.
Latent communication removes explicit natural-language verbalization,
but Latent-MAS still constructs each inter-agent message through a
sequential latent rollout. Thought Blocks target this remaining
construction cost by replacing the rollout with a single non-autoregressive
block-construction pass per inner agent. Under our four-agent setup,
this reduces communication-stage passes from $12$ for Latent-MAS to
$3$ for Thought Blocks; Text-MAS requires hundreds of such passes because
its intermediate messages are decoded autoregressively. Notably, Thought
Blocks and Latent-MAS have the same macro-average text-token count in
Table~\ref{tab:main_results_category_average}, yet Thought Blocks remain
substantially faster. This comparison therefore shows that the efficiency
gain is not explained by shorter textual outputs, but by replacing
sequential latent message construction with one-shot block construction.

\begin{table*}[t]
\centering
\caption{
Main comparison across four LLM backbones and three task categories.
Results are category means. \emph{Improve} compares Ours with Text-MAS:
accuracy is Ours $-$ Text-MAS, efficiency metrics are relative reductions,
and latency is speedup. Bold indicates the best multi-agent result.
}
\label{tab:main_results_category_average}

\begingroup
\setlength{\tabcolsep}{2.2pt}
\renewcommand{\arraystretch}{1}

\scriptsize


\resizebox{\textwidth}{!}{%
\begin{tabular}{
@{}
ll
ccccc
@{\hspace{12pt}}
ccccc
@{}
}
\toprule

&
&
\multicolumn{5}{c}{\textbf{Qwen2.5-7B}}
&
\multicolumn{5}{c}{\textbf{Qwen3-8B}}
\\

\cmidrule(lr){3-7}
\cmidrule(lr){8-12}

\textbf{Task}
&
\textbf{Method}
&
\textbf{Acc.}
&
\textbf{Text Tokens}
&
\textbf{Passes}$^{*}$
&
\textbf{AR Steps}$^{\dagger}$
&
\textbf{Latency}$^{\dagger}$
&
\textbf{Acc.}
&
\textbf{Text Tokens}
&
\textbf{Passes}$^{*}$
&
\textbf{AR Steps}$^{\dagger}$
&
\textbf{Latency}$^{\dagger}$
\\

\midrule


\multirow{5}{*}{Reasoning}
&
Single Text
&
69.06
&
233
&
-
&
53
&
2.18
&
86.45
&
233
&
-
&
50
&
1.75
\\

&
Text-MAS
&
\textbf{74.91}
&
1717
&
483
&
492
&
16.09
&
\textbf{89.75}
&
1546
&
375
&
380
&
16.80
\\

&
Latent-MAS
&
73.85
&
751
&
12
&
30
&
3.85
&
89.62
&
752
&
12
&
16
&
3.99
\\

&
Ours
&
73.77
&
\textbf{737}
&
\textbf{3}
&
\textbf{6}
&
\textbf{0.85}
&
89.24
&
\textbf{738}
&
\textbf{3}
&
\textbf{10}
&
\textbf{1.23}
\\

\cline{2-12}
\noalign{\vskip 1pt}

&
Improve
&
-1.13
&
57.1\%
&
99.4\%
&
98.7\%
&
18.88$\times$
&
-0.51
&
52.3\%
&
99.2\%
&
97.3\%
&
13.66$\times$
\\

\midrule


\multirow{5}{*}{\shortstack{Question\\Answering}}
&
Single Text
&
56.12
&
285
&
-
&
11
&
0.42
&
62.57
&
397
&
-
&
126
&
6.17
\\

&
Text-MAS
&
58.16
&
2298
&
595
&
618
&
20.42
&
70.78
&
2252
&
552
&
574
&
31.31
\\

&
Latent-MAS
&
\textbf{58.90}
&
1117
&
12
&
25
&
3.75
&
\textbf{70.94}
&
1156
&
12
&
\textbf{15}
&
3.37
\\

&
Ours
&
57.89
&
\textbf{1114}
&
\textbf{3}
&
\textbf{7}
&
\textbf{1.05}
&
70.27
&
\textbf{1136}
&
\textbf{3}
&
31
&
\textbf{3.20}
\\

\cline{2-12}
\noalign{\vskip 1pt}

&
Improve
&
-0.27
&
51.5\%
&
99.5\%
&
98.9\%
&
19.37$\times$
&
-0.50
&
49.5\%
&
99.5\%
&
94.6\%
&
9.79$\times$
\\

\midrule


\multirow{5}{*}{Reading}
&
Single Text
&
74.66
&
439
&
-
&
6
&
0.24
&
80.23
&
452
&
-
&
15
&
0.60
\\

&
Text-MAS
&
77.36
&
2502
&
356
&
370
&
12.95
&
\textbf{84.39}
&
2478
&
324
&
346
&
16.36
\\

&
Latent-MAS
&
\textbf{79.31}
&
\textbf{1753}
&
12
&
17
&
3.70
&
84.06
&
1815
&
12
&
14
&
3.95
\\

&
Ours
&
75.63
&
\textbf{1753}
&
\textbf{3}
&
\textbf{7}
&
\textbf{1.61}
&
83.84
&
\textbf{1749}
&
\textbf{3}
&
\textbf{6}
&
\textbf{1.08}
\\

\cline{2-12}
\noalign{\vskip 1pt}

&
Improve
&
-1.73
&
29.9\%
&
99.2\%
&
98.1\%
&
8.02$\times$
&
-0.55
&
29.4\%
&
99.1\%
&
98.3\%
&
15.08$\times$
\\

\bottomrule
\end{tabular}
}

\vspace{3mm}


\resizebox{\textwidth}{!}{%
\begin{tabular}{
@{}
ll
ccccc
@{\hspace{12pt}}
ccccc
@{}
}
\toprule

&
&
\multicolumn{5}{c}{\textbf{Qwen3-14B}}
&
\multicolumn{5}{c}{\textbf{Mistral-7B}}
\\

\cmidrule(lr){3-7}
\cmidrule(lr){8-12}

\textbf{Task}
&
\textbf{Method}
&
\textbf{Acc.}
&
\textbf{Text Tokens}
&
\textbf{Passes}$^{*}$
&
\textbf{AR Steps}$^{\dagger}$
&
\textbf{Latency}$^{\dagger}$
&
\textbf{Acc.}
&
\textbf{Text Tokens}
&
\textbf{Passes}$^{*}$
&
\textbf{AR Steps}$^{\dagger}$
&
\textbf{Latency}$^{\dagger}$
\\

\midrule


\multirow{5}{*}{Reasoning}
&
Single Text
&
85.19
&
222
&
-
&
34
&
1.79
&
59.25
&
270
&
-
&
63
&
2.17
\\

&
Text-MAS
&
89.44
&
1765
&
491
&
496
&
26.14
&
65.27
&
2994
&
704
&
764
&
40.89
\\

&
Latent-MAS
&
\textbf{91.40}
&
\textbf{771}
&
12
&
155
&
17.01
&
\textbf{66.44}
&
\textbf{1405}
&
12
&
581
&
29.94
\\

&
Ours
&
89.99
&
850
&
\textbf{3}
&
\textbf{107}
&
\textbf{9.33}
&
65.72
&
1407
&
\textbf{3}
&
\textbf{286}
&
\textbf{13.12}
\\

\cline{2-12}
\noalign{\vskip 1pt}

&
Improve
&
+0.55
&
51.8\%
&
99.4\%
&
78.4\%
&
2.80$\times$
&
+0.45
&
53.0\%
&
99.6\%
&
62.6\%
&
3.12$\times$
\\

\midrule


\multirow{5}{*}{\shortstack{Question\\Answering}}
&
Single Text
&
69.46
&
318
&
-
&
44
&
2.07
&
39.68
&
447
&
-
&
117
&
4.99
\\

&
Text-MAS
&
\textbf{74.64}
&
2565
&
692
&
711
&
38.34
&
44.89
&
4300
&
993
&
1068
&
62.81
\\

&
Latent-MAS
&
73.91
&
\textbf{1158}
&
12
&
115
&
16.01
&
\textbf{46.85}
&
1831
&
12
&
585
&
38.66
\\

&
Ours
&
74.40
&
1189
&
\textbf{3}
&
\textbf{98}
&
\textbf{8.40}
&
46.28
&
\textbf{1806}
&
\textbf{3}
&
\textbf{275}
&
\textbf{16.46}
\\

\cline{2-12}
\noalign{\vskip 1pt}

&
Improve
&
-0.24
&
53.7\%
&
99.6\%
&
86.1\%
&
4.56$\times$
&
+1.40
&
58.0\%
&
99.7\%
&
74.3\%
&
3.82$\times$
\\

\midrule


\multirow{5}{*}{Reading}
&
Single Text
&
82.78
&
448
&
-
&
7
&
0.44
&
71.28
&
512
&
-
&
32
&
1.03
\\

&
Text-MAS
&
\textbf{86.03}
&
2621
&
412
&
418
&
22.81
&
73.86
&
3666
&
588
&
631
&
34.45
\\

&
Latent-MAS
&
85.72
&
\textbf{1810}
&
12
&
141
&
13.85
&
\textbf{75.34}
&
2487
&
12
&
581
&
35.66
\\

&
Ours
&
85.95
&
1852
&
\textbf{3}
&
\textbf{111}
&
\textbf{11.02}
&
74.69
&
\textbf{2475}
&
\textbf{3}
&
\textbf{272}
&
\textbf{15.39}
\\

\cline{2-12}
\noalign{\vskip 1pt}

&
Improve
&
-0.08
&
29.3\%
&
99.3\%
&
73.5\%
&
2.07$\times$
&
+0.83
&
32.5\%
&
99.5\%
&
56.9\%
&
2.24$\times$
\\

\bottomrule
\end{tabular}
}

\vspace{1.5mm}

\begin{minipage}{0.98\textwidth}
\scriptsize
\textit{Note.}
$^{*}$ Communication-stage metric, measured over intermediate-agent message construction.
$^{\dagger}$ End-to-end metric, measured from input to final output; latency is in seconds.
\end{minipage}

\endgroup

\vspace{-4mm}
\end{table*}




\subsection{Ablation Study}
\label{app:ablation_study}

We ablate two design choices of Thought Blocks on Qwen3-8B with
$B=4$. First, we test whether the block positions require
writer-generated content by replacing the block embeddings $E_i$ with
zero or random vectors while leaving the rest of the pipeline
unchanged. Second, we evaluate the factorized continuation state
$\mathcal{S}_i=(R_i,C_i)$ by retaining only $R_i$ or only $C_i$.

As shown in Table~\ref{tab:ablation_accuracy}, replacing the
writer-generated content of $E_i$ consistently degrades performance
across all eleven benchmark configurations. The average accuracy drops
from $79.64$ for the full method to $73.44$ with
\textsc{Zero-$E$} ($-6.20$ points) and to $71.06$ with
\textsc{Random-$E$} ($-8.58$ points). Moreover,
\textsc{Random-$E$} underperforms \textsc{Zero-$E$} on ten of the eleven
benchmarks, with an additional average drop of $2.38$ points. These
results rule out the possibility that the gains of Thought Blocks arise
merely from introducing additional latent positions. Instead, the
content constructed at these positions is functionally important, and
replacing it with arbitrary latent values can interfere with downstream
computation.
\begin{table*}[htbp]
\vspace{-4mm}
  \centering
  \caption{Ablation study of our method and its variants on the datasets. Accuracy (\%) is reported.}
  \label{tab:ablation_accuracy}

  \resizebox{\textwidth}{!}{
  \begin{tabular}{@{}lccccccccccc@{}}
    \toprule
    \textbf{Method}
    & \textbf{ARC-C}
    & \textbf{ARC-E}
    & \textbf{BoolQ}
    & \textbf{GPQA}
    & \textbf{GSM8K}
    & \textbf{MedQA(US)}
    & \textbf{MedQA(TW)}
    & \textbf{MedQA(ML)}
    & \textbf{MMLU}
    & \textbf{OpenBookQA}
    & \textbf{QuAIL}
    \\
    \midrule

    Ours
    & 91.76
    & 96.91
    & 90.25
    & 43.89
    & 82.05
    & 71.16
    & 73.41
    & 80.12
    & 82.79
    & 86.23
    & 77.44
    \\

    Zero $E$
    & 83.00
    & 84.96
    & 83.89
    & 32.54
    & 70.40
    & 68.01
    & 71.72
    & 79.22
    & 80.09
    & 82.71
    & 71.25
    \\

    Random $E$
    & 80.70
    & 83.25
    & 79.52
    & 30.37
    & 70.56
    & 66.35
    & 68.81
    & 76.43
    & 77.61
    & 78.74
    & 69.28
    \\

    \midrule

    $R-only$
    & 79.11
    & 78.63
    & 71.39
    & 15.31
    & 58.94
    & 54.66
    & 57.98
    & 68.25
    & 65.72
    & 64.43
    & 59.67
    \\

    $C-only$
    & 82.63
    & 82.94
    & 76.08
    & 19.15
    & 66.36
    & 60.32
    & 64.57
    & 75.61
    & 72.43
    & 73.65
    & 64.98
    \\

    \bottomrule
  \end{tabular}
  }
\vspace{-2mm}
\end{table*}

The state-factorization ablations show that both components are
important for recovering the performance of the full method.
Retaining only $R_i$ reduces the average accuracy to $61.28$
($-18.36$ points), while retaining only $C_i$ yields $67.16$
($-12.48$ points). The $C$-only variant outperforms the $R$-only
variant on all eleven benchmarks, by $5.88$ points on average,
suggesting that the addressable contextual memory carried by $C_i$ is
particularly valuable when each component is considered in isolation.
Nevertheless, both single-component variants remain substantially
below the full method. This is consistent with the intended
factorization: $R_i$ provides the continuation signal used to construct
the block positions, whereas $C_i$ preserves addressable prefix memory
that those positions can retrieve during rendering. Full performance
therefore depends on using the two components jointly.

\subsection{Effect of Thought Block Size}

We investigate how the Thought Block size $B$ affects the trade-off between communication capacity and efficiency on Qwen3-8B. As shown in Table~\ref{tab:block_size_avg}, even $B=1$ retains substantial task utility, achieving $78.05\%$ category-macro average accuracy. Increasing $B$ to $4$ improves accuracy to $81.12\%$, while further increasing it to $32$ yields only modest additional gains, reaching $82.59\%$.

Meanwhile, latency remains nearly constant because all block positions are rendered in a single backbone forward pass, whereas KV-cache load grows with $B$. The strong $B=1$ result is also consistent with our design: each slot is rendered into a full layer-wise KV continuation position conditioned on the compressed context bank. Overall, $B=4$ provides a favorable balance between accuracy and memory cost.

\begin{table*}[htbp]
\vspace{-4mm}
\centering
\caption{Effect of Thought Block size on Qwen3-8B.}
\label{tab:block_size_avg}

\setlength{\tabcolsep}{4pt}
\resizebox{0.65\textwidth}{!}{
\begin{tabular}{clrrrrrrr}
\toprule
\textbf{Task Category} & \textbf{Metric} &\textbf{B=1}
& \textbf{B=4} & \textbf{B=8} & \textbf{B=12} 
& \textbf{B=16} & \textbf{B=24} & \textbf{B=32} \\
\midrule

\multirow{3}{*}{Reasoning}
& Accuracy 
&87.14 & 89.24 & 89.46 & 89.69 & 89.35 & 89.34 & \textbf{90.02} \\

& Latency (s)
&\textbf{1.13} & 1.23 & 1.26 & 1.27 & 1.21 & 1.25 & 1.29 \\

& KV Load (MB) 
&4.66 & 5.71 & 7.12 & 8.52 & 9.93 & 12.74 & 15.55 \\

\midrule

\multirow{3}{*}{Question Answering}
& Accuracy 
&65.49 & 70.27 & 71.24 & 71.23 & 71.66 & 71.82 & \textbf{72.21} \\

& Latency (s)
&\textbf{3.12} & 3.20 & 3.23 & 3.27 & 3.23 & 3.20 & 3.34 \\

& KV Load (MB)
&7.37 & 8.42 & 9.83 & 11.23 & 12.64 & 15.45 & 18.26 \\

\midrule

\multirow{3}{*}{Reading}
& Accuracy 
&81.53 & 83.84 & 84.58 & 84.41 & 84.72 & 85.04 & \textbf{85.55} \\

& Latency (s)
&\textbf{0.94} & 1.08 & 1.12 & 1.10 & 1.10 & 1.13 & 1.14 \\

& KV Load (MB)
&12.39 & 13.44 & 14.84 & 16.24 & 17.66 & 20.47 & 23.28 \\

\bottomrule
\end{tabular}
}
\vspace{-2mm}
\end{table*} 

These results show diminishing returns from increasing the Thought Block size: even $B=4$ achieves most of the accuracy obtained with larger blocks, while KV-cache load increases approximately linearly with $B$. Thus, within the tested range, $B$ mainly controls the accuracy--memory trade-off rather than acting as a highly performance-sensitive hyperparameter.

\subsection{Decoding-Time KV-Cache Load}

Table~\ref{tab:load_kv} reports decoding-time KV-cache load. Thought Blocks load 8.35 MB on average, compared with 58.29 MB for Latent-MAS and 71.13 MB for Text-MAS, reducing loaded KV-cache volume by
85.7\% and 88.3\%, respectively. This shows that the efficiency gain extends
beyond token count and sequential forward steps to substantially lower
decoding-time KV-cache load.
\begin{table*}[htbp]
\vspace{-4mm}
  \centering
  \caption{
Loaded KV-cache size (MB) during decoding on Qwen3-8B.
}
  \label{tab:load_kv}

  \resizebox{\textwidth}{!}{
  \begin{tabular}{@{}lccccccccccc@{}}
    \toprule
    \textbf{Method}
    & \textbf{ARC-C}
    & \textbf{ARC-E}
    & \textbf{BoolQ}
    & \textbf{GPQA}
    & \textbf{GSM8K}
    & \textbf{MedQA(ML)}
    & \textbf{MedQA(TW)}
    & \textbf{MedQA(US)}
    & \textbf{MMLU}
    & \textbf{OpenBookQA}
    & \textbf{QuAIL}\\
    \midrule

    Ours
    & \textbf{6.25}
    & \textbf{5.66}
    & \textbf{8.61}
    & \textbf{10.13}
    & \textbf{5.61}
    & \textbf{6.78}
    & \textbf{7.28}
    & \textbf{10.76}
    & \textbf{7.17}
    & \textbf{5.33}
    & \textbf{18.26}\\

    Latent-MAS
    & 41.96
    & 39.95
    & 59.48
    & 71.42
    & 38.99
    & 47.08
    & 51.14
    & 75.68
    & 52.87
    & 36.42
    & 126.18\\

    Text-MAS
    & 58.02
    & 55.49
    & 64.83
    & 93.86
    & 50.43
    & 67.54
    & 75.49
    & 89.27
    & 65.75
    & 52.79
    & 108.91\\

    \bottomrule
  \end{tabular}
  }
\vspace{-4mm}
\end{table*}

\section{Discussion and Limitations}

Our results support a distinction between message representation and message construction. Thought Blocks show that useful continuation states can be constructed directly from the sender state without executing an autoregressive latent rollout at inference, while preserving comparable downstream utility across multiple backbones and tasks. This suggests that autoregression is one mechanism for reaching a useful latent state, rather than an intrinsic requirement of latent communication. The results nevertheless reflect an accuracy--efficiency trade-off: performance varies across tasks and models, and larger blocks provide diminishing accuracy gains at increasing KV-memory cost.

Our study focuses on establishing whether autoregressive latent message construction can be compiled into a direct continuation-state mapping. The current training pipeline uses autoregressive teacher rollouts to provide offline state-level supervision, but this sequential computation is removed entirely at inference. Consequently, the additional training cost is a one-time overhead whose importance depends on deployment scale; our amortization analysis in Appendix~\ref{app:amortized_efficiency} suggests that it becomes relatively small in sufficiently high-volume multi-agent workloads. We evaluate this principle in controlled sequential, shared-backbone systems in order to isolate message-construction effects. 
Extending direct state construction to heterogeneous backbones, richer communication graphs, and long-horizon tool-using agents is an important direction for future work. Finally, latent KV messages are not directly human-readable, so applications requiring auditable intermediate artifacts may benefit from hybrid protocols that selectively externalize text while using one-shot latent communication on internal, high-frequency communication edges. More broadly, we view Thought Blocks not as a universal replacement for text or sequential reasoning, but as an efficient communication primitive that can be deployed selectively when intermediate trajectories need not themselves be exposed.
\clearpage
\bibliographystyle{iclr2027_conference}
\bibliography{references}

\clearpage
\appendix
\section*{Appendix}
\setcounter{table}{0}
\setcounter{figure}{0}
\renewcommand{\thetable}{\thesection .\arabic{table}}
\renewcommand{\thefigure}{\thesection .\arabic{figure}}
\section{Method Overview}
\begin{figure}[htbp]
  \centering
  \includegraphics[width=1\textwidth]{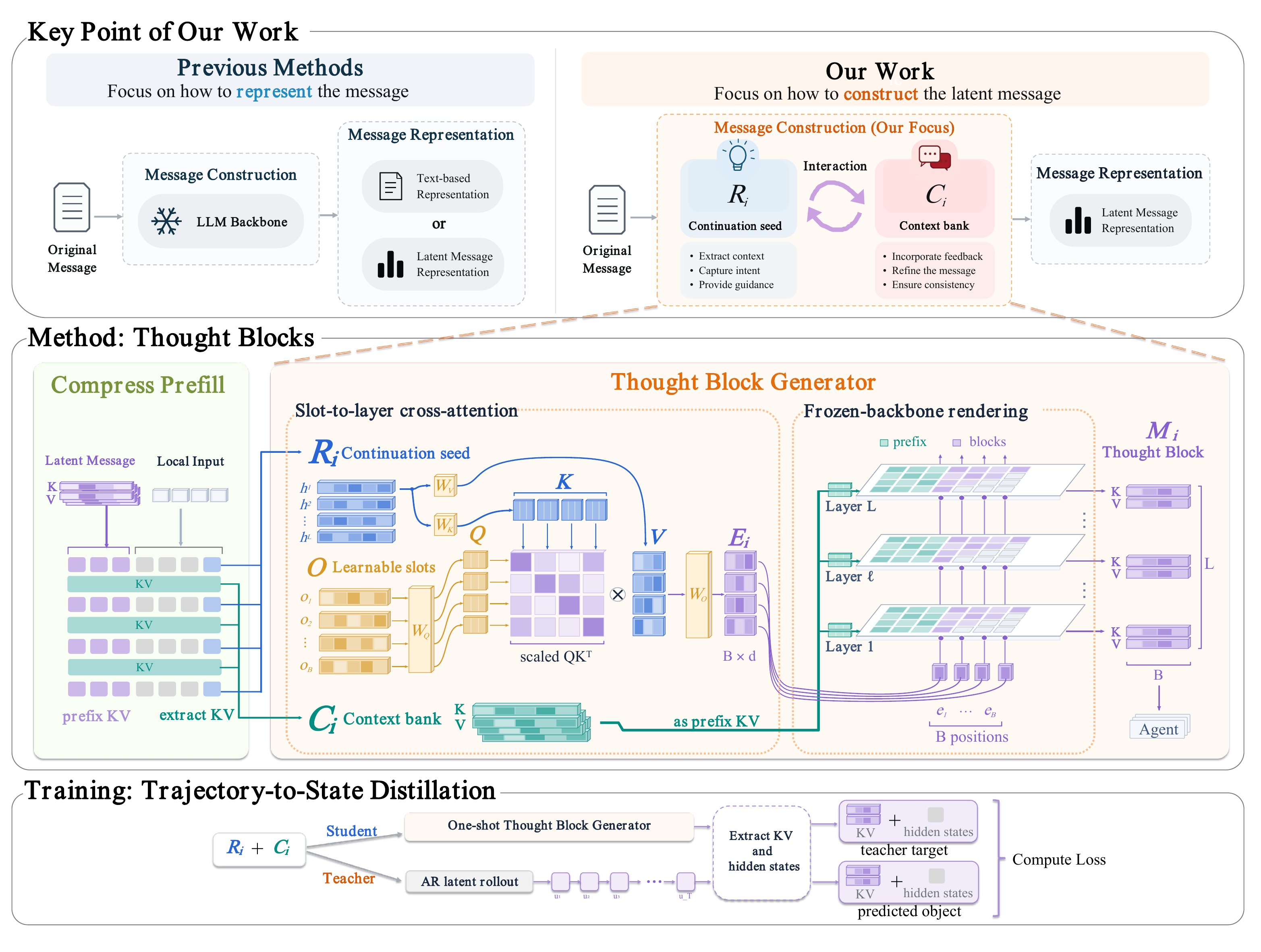}
  \caption{Overview of our Thought Block framework for latent multi-agent communication. Each agent first encodes its local context together with the compressed communication history, where earlier interactions are retained through Beacon KV states while the most recent context preserves its raw-token states. During training, a frozen Beacon teacher autoregressively constructs a latent continuation trajectory, from which a fixed number of model-native continuation states are extracted to form the supervision target. The Thought Block Writer is then trained to predict these continuation-memory states directly from the pre-trajectory representation. At inference time, the teacher trajectory is removed entirely: the Writer generates the Thought Block in a single forward pass, and the resulting latent states are directly propagated to subsequent agents through the KV cache without decoding intermediate textual messages. This design preserves model-native latent information while substantially reducing the autoregressive communication overhead of conventional text-based multi-agent systems.}
  \label{fig:main_method_pic}
\end{figure}

\section{Experiment Setups}
\subsection{Training Thought Blocks}
\label{app:thought_block_training}
Training follows the same trajectory-to-state view introduced in Sec.~\ref{sec:trajectory_distillation}. Stage I trains the Activation Beacon mechanism used to instantiate the compressed context bank while keeping the language-model backbone fixed. Stage II then performs trajectory-to-state distillation. Starting from the factorized continuation state $\mathcal{S}_i=(R_i,C_i)$, a frozen autoregressive latent teacher executes the trajectory $\tau_i$ and a state-extraction operator $\Pi$ converts that trajectory into a fixed-size teacher continuation-memory target $\mathcal{M}_i^\star$. The one-shot Thought Block writer receives only $(R_i,C_i)$ and is trained to reconstruct this target directly. The teacher trajectory is therefore used only to construct supervision during training and is removed entirely at inference time.

\subsubsection{Stage I: Beacon Training}

We instantiate the context bank $C_i$ with Activation Beacon~\citep{zhang2024activation}. Let $\theta$ denote the backbone parameters and let $\beta$ denote the trainable Beacon mechanism, including the Beacon embedding and Beacon-specific projections. Given a training corpus $\mathcal{D}$, we optimize only $\beta$ with the Activation Beacon objective,
\begin{equation}
\beta^\star
=
\arg\min_{\beta}
\mathbb{E}_{x\sim\mathcal{D}}
\left[
\mathcal{L}_{\mathrm{AB}}
\bigl(\mathrm{Beacon}_{\beta}(x);\theta\bigr)
\right].
\end{equation}
The language-model weights are not updated in this stage. We use $\theta^\star$ to denote this frozen backbone in the remainder of the paper and freeze the trained Beacon mechanism $\beta^\star$ as well. Consequently, Stage II optimizes only the writer parameters $\phi$, including learnable slots $O$.

\subsubsection{Stage II-A: Teacher State Construction}

For each training example $x_i$, we first run the same Beacon-compressed prefill used by the main method to obtain the pre-trajectory continuation state
\begin{equation}
\mathcal{S}_i=(R_i,C_i),
\end{equation}
where $R_i$ is the layer-wise continuation seed at the final prefill position and $C_i$ is the compressed, addressable prefix KV memory defined in Sec.~\ref{sec:factorized_state}.

The frozen backbone then performs autoregressive latent decoding from this state. Following autoregressive latent communication~\citep{zou2025latent}, the last-layer teacher hidden state is mapped back to the model input-embedding space before being used as the next continuous input. Under the row-vector convention, we use the fixed alignment operator
\begin{equation}
W_a
=
W_{\mathrm{out}}^{\dagger}W_{\mathrm{in}}
\in \mathbb{R}^{d\times d},
\end{equation}
where $W_{\mathrm{in}},W_{\mathrm{out}}\in\mathbb{R}^{|\mathcal{V}|\times d}$ are the backbone input- and output-embedding matrices and $W_{\mathrm{out}}^{\dagger}$ is the Moore--Penrose pseudoinverse. If $h^{\mathrm{teach}}_{i,t}$ is the last-layer teacher hidden state at latent step $t$, the continuous latent state used by the next step is
\begin{equation}
u_{i,t}=h^{\mathrm{teach}}_{i,t}W_a.
\end{equation}
The resulting teacher trajectory is exactly the trajectory used in the main text,
\begin{equation}
\tau_i
=
[u_{i,1},\ldots,u_{i,T_{\mathrm{lat}}}]
=
\mathrm{ARDecode}_{\theta^\star}(R_i;C_i).
\end{equation}

To instantiate the state-extraction operator $\Pi$ in Eq.~\ref{eq:teacher_state}, we insert decode-stage Beacons along the teacher rollout and retain the states at the $B$ valid Beacon positions. The extracted teacher continuation-memory target is
\begin{equation}
\mathcal{M}_i^\star
=
\Pi(\tau_i)
=
\left\{
\left(
K_{i,j}^{\ell,\star},
V_{i,j}^{\ell,\star}
\right)
\right\}_{\substack{\ell=1,\ldots,L\\j=1,\ldots,B}}.
\label{eq:app_teacher_kv_target}
\end{equation}
At the same retained positions, we additionally record the corresponding last-layer hidden states
\begin{equation}
Z_i^\star
=
\left\{z_{i,j}^\star\right\}_{j=1}^{B},
\label{eq:app_teacher_aux_target}
\end{equation}
which are used only for the auxiliary training loss and are not part of the communicated message. Thus, each collected supervision record contains $(R_i,C_i,\mathcal{M}_i^\star,Z_i^\star)$. This construction makes the distillation target model-native: the writer learns to reconstruct continuation states actually reached by the frozen autoregressive teacher rather than arbitrary latent tensors.

\subsubsection{Stage II-B: Thought Block Writer Training}

Given $(R_i,C_i)$, the writer first performs the same parallel continuation expansion as in Sec.~\ref{sec:thought_block_writer},
\begin{equation}
E_i=g_\phi(O,R_i),
\end{equation}
and then renders the $B$ virtual continuation positions through the frozen backbone conditioned on $C_i$,
\begin{equation}
\widehat{\mathcal{M}}_i
=
\mathrm{Render}_{\theta^\star}(E_i;C_i).
\end{equation}
The same forward pass also produces the last-layer hidden states used by the auxiliary objective. We denote them by
\begin{equation}
\widehat{Z}_i
=
\left\{\hat z_{i,j}\right\}_{j=1}^{B},
\qquad
\widehat{\mathcal{M}}_i
=
\left\{
\left(
\widehat{K}_{i,j}^{\ell},
\widehat{V}_{i,j}^{\ell}
\right)
\right\}_{\substack{\ell=1,\ldots,L\\j=1,\ldots,B}}.
\end{equation}

The primary objective is the same KV-state reconstruction loss defined in Eq.~\ref{eq:kv_loss},
\begin{equation}
\mathcal{L}_{\mathrm{KV}}
=
\frac{1}{BL}
\sum_{\ell=1}^{L}
\sum_{j=1}^{B}
\left(
\left\|
\widehat{K}_{i,j}^{\ell}-K_{i,j}^{\ell,\star}
\right\|_F^2
+
\left\|
\widehat{V}_{i,j}^{\ell}-V_{i,j}^{\ell,\star}
\right\|_F^2
\right).
\end{equation}
The auxiliary hidden-state alignment objective deferred from the main text is
\begin{equation}
\mathcal{L}_{\mathrm{aux}}
=
\frac{1}{B}
\sum_{j=1}^{B}
\left\|
\hat z_{i,j}-z_{i,j}^{\star}
\right\|_2^2.
\label{eq:app_aux_loss}
\end{equation}
The complete trajectory-to-state distillation objective is therefore
\begin{equation}
\mathcal{L}_{\mathrm{TSD}}
=
\lambda_{\mathrm{KV}}\mathcal{L}_{\mathrm{KV}}
+
\lambda_{\mathrm{aux}}\mathcal{L}_{\mathrm{aux}},
\end{equation}
which matches Eq.~\ref{eq:tsd_loss}. Only $\phi$ is updated in Stage II; $\theta^\star$ and $\beta^\star$ remain frozen. At inference time, neither the autoregressive teacher trajectory $\tau_i$ nor the auxiliary target $Z_i^\star$ is needed: the writer directly applies $(R_i,C_i)\mapsto\mathcal{M}_i$ in one forward pass.


\clearpage
\subsection{Prompt Templates}
\label{app:prompt_templates}
\subsubsection{Text-MAS Prompt Templates}
\label{app:text_prompts}

\begin{agentprompt}{Planner Agent Prompt}

You are a Planner Agent. Given an input question, design a clear, step-by-step plan for how to solve the question.

Your outlined plan should be concise with a few bulletpoints for each step. Do not produce the final answer.

Your plan will be passed as plain text to later agents.

\texttt{\{question\}}

Now output your plan to solve the question below:

\end{agentprompt}

\begin{agentprompt}{Critic Agent Prompt}

You are a Critic Agent to evaluate the correctness of the input plan for the given question and provide helpful feedback for improving the plan.

The plan information is provided in plain text from the previous Planner agent. Review the plan and question and output:

(1) original plan contents

(2) constructive feedback on the original plan.

\texttt{\{question\}}

Now, output your response below:

\end{agentprompt}

\begin{agentprompt}{Refiner Agent Prompt}

You are a Refiner Agent to provide a refined step-by-step plan for solving the given question.

You are provided with:

(1) text-format information: a previous plan with feedback from earlier agents and the input question you need to solve.

\texttt{\{question\}}

Based on the input, write a refined and improved plan to solve the question. Make sure your output plan is correct and concise.

Now, output your refined plan below:

\end{agentprompt}

\begin{agentprompt}{Judger Agent Prompt}

You are a helpful assistant. You are provided with textual information from previous agents for reference and a target question to solve.

The previous agents' text might contain irrelevant contents. Ignore it if it is not helpful for solving the target question.

You must reason step-by-step to solve the provided Target Question without outputting other irrelevant information.

\texttt{\{question\}}

Now, output your final answer and just output one answer:

\end{agentprompt}

\clearpage
\subsubsection{Latent-MAS Prompt Templates}
\label{app:latent_prompts}

\begin{agentprompt}{Planner Agent Prompt}

You are a Planner Agent. Given an input question, design a clear, step-by-step plan for how to solve the question.

Your outlined plan should be concise with a few bulletpoints for each step. Do not produce the final answer.

\texttt{\{question\}}

Now output your plan to solve the question below:

\end{agentprompt}

\begin{agentprompt}{Critic Agent Prompt}

You are a Critic Agent to evaluate the correctness of the input plan for the given question and provide helpful feedback for improving the plan.

The plan information is provided in latent KV representation format. Review the plan and question and output:

(1) original plan contents

(2) constructive feedback on the original plan.

\texttt{\{question\}}

Now, output your response below:

\end{agentprompt}

\begin{agentprompt}{Refiner Agent Prompt}

You are a Refiner Agent to provide a refined step-by-step plan for solving the given question.

You are provided with:

(1) latent-format information: a previous plan with feedback and the input question you need to solve.

\texttt{\{question\}}

Based on the input, write a refined and improved plan to solve the question. Make sure your output plan is correct and concise.

Now, output your refined plan below:

\end{agentprompt}

\begin{agentprompt}{Judger Agent Prompt}

You are a helpful assistant. You are provided with latent information for reference and a target question to solve.

The latent information might contain irrelevant contents. Ignore it if it is not helpful for solving the target question.

You must reason step-by-step to solve the provided Target Question without outputting other irrelevant information.

\texttt{\{question\}}

Now, output your final answer and just output one answer:

\end{agentprompt}

\clearpage
\subsubsection{Thought Blocks Prompt Templates}
\label{app:ours_prompts}

\begin{agentprompt}{Planner Agent Prompt}

You are a Planner Agent. Given an input question, design a clear, step-by-step plan for how to solve the question.

Your outlined plan should be concise with a few bulletpoints for each step. Do not produce the final answer.

Your plan will be compressed into beacon latent memory and passed to later agents.

\texttt{\{question\}}

Now output your plan to solve the question below:

\end{agentprompt}

\begin{agentprompt}{Critic Agent Prompt}

You are a Critic Agent to evaluate the correctness of the input plan for the given question and provide helpful feedback for improving the plan.

The plan information is provided in beacon-compressed latent representation format. Review the plan and question and output:

(1) original plan contents

(2) constructive feedback on the original plan.

\texttt{\{question\}}

Now, output your response below:

\end{agentprompt}

\begin{agentprompt}{Refiner Agent Prompt}

You are a Refiner Agent to provide a refined step-by-step plan for solving the given question.

You are provided with:

(1) beacon-compressed latent information: a previous plan with feedback and the input question you need to solve.

\texttt{\{question\}}

Based on the input, write a refined and improved plan to solve the question. Make sure your output plan is correct and concise.

Now, output your refined plan below:

\end{agentprompt}

\begin{agentprompt}{Judger Agent Prompt}

You are a helpful assistant. You are provided with beacon-compressed latent information from previous agents for reference and a target question to solve.

The beacon-compressed latent information might contain irrelevant or lossy contents. Ignore it if it is not helpful for solving the target question.

You must reason step-by-step to solve the provided Target Question without outputting other irrelevant information.

\texttt{\{question\}}

Now, output your final answer and just output one answer:

\end{agentprompt}

\clearpage
\subsection{Full Experiment Results}
\begin{table*}[htbp]
\centering
\small
\setlength{\tabcolsep}{4pt}
\renewcommand{\arraystretch}{0.92}
\caption{Main performance comparison on all models across benchmarks. For accuracy, Improve is computed as Ours $-$ Text-MAS. For efficiency metrics, Improve reports the reduction or speedup of Ours relative to Text-MAS. The best and second-best results among Ours, Latent-MAS, and Text-MAS are highlighted in bold and underlined, respectively. For Improve rows with an accuracy difference greater than $-1.5$ points. (Part 1/6: ARC-C, ARC-E.)}
\label{tab:full_main_results}
\resizebox{\textwidth}{!}{%
\begin{tabular}{@{}p{1.5cm}p{1.9cm}lccccc@{}}
\toprule
\textbf{Dataset} & \textbf{Model} & \textbf{Method}
& \textbf{Acc.}
& \textbf{Text Tokens}
& \textbf{Comm. Passes}
& \textbf{AR Steps}
& \textbf{Latency}\\
\midrule
\multirow{20}{*}{\centering ARC-C}
& \multirow{5}{*}{Qwen2.5-7B}
& Ours & 82.38 & \textbf{784} & 3 & \textbf{4} & \textbf{0.68} \\
& & Latent-MAS & \underline{83.39} & \underline{788} & 12 & \underline{17} & \underline{3.39} \\
& & Text-MAS & \textbf{83.96} & 1757 & 479& 482 & 13.36 \\
& & Single & 81.19 & 199 & - & 5 & 0.09 \\
& & Improve
& -1.58
& 55.4\%
& 99.4\%
& 99.6\%
& 19.6$\times$ \\
\cmidrule(rl){2-8}

& \multirow{5}{*}{Qwen3-8B}
& Ours & \underline{91.76} & \textbf{769} & 3 & \textbf{4} & \textbf{0.61} \\
& & Latent-MAS & 90.19 & \underline{800} & 12 & \underline{14} & \underline{3.52} \\
& & Text-MAS & \textbf{92.06} & 1653 & 417& 419 & 15.65 \\
& & Single & 89.10 & 210 & - & 10 & 0.36 \\
& & \cellcolor{green!12}Improve
& \cellcolor{green!12}-0.30
& \cellcolor{green!12}53.5\%
& \cellcolor{green!12}99.3\%
& \cellcolor{green!12}99.5\%
& \cellcolor{green!12}25.7$\times$ \\
\cmidrule(rl){2-8}

& \multirow{5}{*}{Qwen3-14B}
& Ours & 90.14 & \underline{879} & 3 & \textbf{93} & \textbf{7.87} \\
& & Latent-MAS & \textbf{91.24} & \textbf{821} & 12 & \underline{107} & \underline{17.77} \\
& & Text-MAS & \underline{90.48} & 1951 & 553& 557 & 29.44 \\
& & Single & 89.13 & 214 & - & 11 & 0.60 \\
& & \cellcolor{green!12}Improve
& \cellcolor{green!12}-0.34
& \cellcolor{green!12}54.9\%
& \cellcolor{green!12}99.5\%
& \cellcolor{green!12}83.3\%
& \cellcolor{green!12}3.7$\times$ \\

\cmidrule(rl){2-8}

& \multirow{5}{*}{Mistral-7B}
& Ours & \underline{71.84} & 1456.78 & 3 & 292 & 17 \\
& & Latent-MAS & \textbf{72.27} & \underline{1480} & 12 & \underline{585} & \underline{38} \\
& & Text-MAS & 71.76 & \textbf{3278} & 806& \textbf{862} & \textbf{52} \\
& & Single & 69.03 & 271 & - & 49 & 1 \\
& & \cellcolor{green!12}Improve
& \cellcolor{green!12}+0.08
& \cellcolor{green!12}55.56%
& \cellcolor{green!12}99.6\%
& \cellcolor{green!12}66.13%
& \cellcolor{green!12}3$\times$ \\
\midrule
\multirow{20}{*}{\centering ARC-E}
& \multirow{5}{*}{Qwen2.5-7B}
& Ours & 91.07 & \textbf{746} & 3 & \textbf{4} & \textbf{0.98} \\
& & Latent-MAS & \underline{91.33} & \underline{748} & 12 & \underline{16} & \underline{2.45} \\
& & Text-MAS & \textbf{92.21} & 1644 & 444& 449 & 15.40 \\
& & Single & 91.81 & 187 & - & 7 & 0.13 \\
& & \cellcolor{green!12}Improve
& \cellcolor{green!12}-1.14
& \cellcolor{green!12}54.6\%
& \cellcolor{green!12}99.3\%
& \cellcolor{green!12}99.6\%
& \cellcolor{green!12}15.7$\times$ \\
\cmidrule(rl){2-8}

& \multirow{5}{*}{Qwen3-8B}
& Ours & 96.91 & \textbf{741} & 3 & \textbf{4} & \textbf{0.69} \\
& & Latent-MAS & \underline{97.14} & \underline{762} & 12 & \underline{14} & \underline{3.97} \\
& & Text-MAS & \textbf{97.35} & 1583 & 399& 404 & 17.11 \\
& & Single & 94.17 & 203 & - & 18 & 0.62 \\
& & \cellcolor{green!12}Improve
& \cellcolor{green!12}-0.44
& \cellcolor{green!12}53.2\%
& \cellcolor{green!12}99.2\%
& \cellcolor{green!12}99.5\%
& \cellcolor{green!12}24.8$\times$ \\
\cmidrule(rl){2-8}

& \multirow{5}{*}{Qwen3-14B}
& Ours & \textbf{97.62} & \underline{824} & 3 & \textbf{88} & \textbf{11.59} \\
& & Latent-MAS & 96.39 & \textbf{774} & 12 & \underline{181} & \underline{17.52} \\
& & Text-MAS & \underline{97.12} & 1855 & 529& 532 & 28.05 \\
& & Single & 93.28 & 198 & - & 7 & 0.39 \\
& & \cellcolor{green!12}Improve
& \cellcolor{green!12}+0.50
& \cellcolor{green!12}55.6\%
& \cellcolor{green!12}99.4\%
& \cellcolor{green!12}83.5\%
& \cellcolor{green!12}2.4$\times$ \\

\cmidrule(rl){2-8}

& \multirow{5}{*}{Mistral-7B}
& Ours & \underline{82.14} & \underline{1446.18} & 3 & 288 & 10 \\
& & Latent-MAS & 81.83 & 1414 & 12 & \underline{580} & \underline{19} \\
& & Text-MAS & \textbf{82.21} & \textbf{2992} & 651& \textbf{751} & \textbf{23} \\
& & Single & 80.04 & 246 & - & 35 & 1 \\
& & \cellcolor{green!12}Improve
& \cellcolor{green!12}-0.07
& \cellcolor{green!12}51.66%
& \cellcolor{green!12}99.5\%
& \cellcolor{green!12}61.65%
& \cellcolor{green!12}2$\times$ \\
\bottomrule
\end{tabular}%
}
\end{table*}


\begin{table*}[t]
\centering
\small
\setlength{\tabcolsep}{4pt}
\renewcommand{\arraystretch}{0.92}
\caption{Main performance comparison on all models across benchmarks (continued; Part 2/6: BoolQ, GPQA).}
\label{tab:full_main_results_part2}
\resizebox{\textwidth}{!}{%
\begin{tabular}{@{}p{1.5cm}p{1.9cm}lccccc@{}}
\toprule
\textbf{Dataset} & \textbf{Model} & \textbf{Method}
& \textbf{Acc.}
& \textbf{Text Tokens}
& \textbf{Comm. Passes}
& \textbf{AR Steps}
& \textbf{Latency}\\
\midrule
\multirow{20}{*}{\centering BoolQ}
& \multirow{5}{*}{Qwen2.5-7B}
& Ours & 84.19 & \textbf{1118} & 3 & \textbf{5} & \textbf{1.49} \\
& & Latent-MAS & \textbf{86.76} & \underline{1119} & 12 & \underline{16} & \underline{3.35} \\
& & Text-MAS & \underline{85.85} & 1938 & 400& 404 & 13.92 \\
& & Single & 83.73 & 280 & - & 6 & 0.16 \\
& & Improve
& -1.66
& 42.3\%
& 99.2\%
& 99.3\%
& 9.3$\times$ \\
\cmidrule(rl){2-8}

& \multirow{5}{*}{Qwen3-8B}
& Ours & \underline{90.25} & \textbf{1121} & 3 & \textbf{6} & \textbf{0.68} \\
& & Latent-MAS & \textbf{90.76} & \underline{1133} & 12 & \underline{15} & \underline{2.56} \\
& & Text-MAS & 90.12 & 1849 & 337& 342 & 20.36 \\
& & Single & 87.49 & 292 & - & 13 & 0.52 \\
& & \cellcolor{green!12}Improve
& \cellcolor{green!12}+0.13
& \cellcolor{green!12}39.4\%
& \cellcolor{green!12}99.1\%
& \cellcolor{green!12}98.8\%
& \cellcolor{green!12}29.9$\times$ \\
\cmidrule(rl){2-8}

& \multirow{5}{*}{Qwen3-14B}
& Ours & 91.33 & \underline{1230} & 3 & \underline{121} & \underline{9.33} \\
& & Latent-MAS & \underline{91.97} & \textbf{1153} & 12 & \textbf{116} & \textbf{8.64} \\
& & Text-MAS & \textbf{92.14} & 1964 & 394& 398 & 21.25 \\
& & Single & 88.29 & 288 & - & 6 & 0.29 \\
& & \cellcolor{green!12}Improve & \cellcolor{green!12}-0.81 & \cellcolor{green!12}37.4\% & \cellcolor{green!12}99.2\% & \cellcolor{green!12}69.6\% & \cellcolor{green!12}2.3$\times$ \\

\cmidrule(rl){2-8}

& \multirow{5}{*}{Mistral-7B}
& Ours & \textbf{87.96} & 1829.51 & 3 & 248 & 17 \\
& & Latent-MAS & \underline{87.75} & \underline{1837} & 12 & \underline{583} & \textbf{36} \\
& & Text-MAS & 86.58 & \textbf{3017} & 593& \textbf{642} & \underline{31} \\
& & Single & 83.30 & 352 & - & 40 & 1 \\
& & \cellcolor{green!12}Improve
& \cellcolor{green!12}+1.38
& \cellcolor{green!12}39.35%
& \cellcolor{green!12}99.5\%
& \cellcolor{green!12}61.37%
& \cellcolor{green!12}2$\times$ \\
\midrule
\multirow{20}{*}{\centering GPQA}
& \multirow{5}{*}{Qwen2.5-7B}
& Ours & \underline{36.65} & \underline{1354} & 3 & \textbf{13} & \textbf{1.50} \\
& & Latent-MAS & \textbf{37.25} & \textbf{1337} & 12 & \underline{34} & \underline{4.46} \\
& & Text-MAS & 35.59 & 2719 & 678& 739 & 25.60 \\
& & Single & 36.39 & 350 & - & 15 & 0.53 \\
& & \cellcolor{green!12}Improve
& \cellcolor{green!12}+1.06
& \cellcolor{green!12}50.2\%
& \cellcolor{green!12}99.6\%
& \cellcolor{green!12}98.2\%
& \cellcolor{green!12}17.1$\times$ \\
\cmidrule(rl){2-8}

& \multirow{5}{*}{Qwen3-8B}
& Ours & \underline{43.89} & \textbf{1414} & 3 & \underline{106} & \underline{7.19} \\
& & Latent-MAS & 43.46 & \underline{1461} & 12 & \textbf{16} & \textbf{2.83} \\
& & Text-MAS & \textbf{44.14} & 2729 & 672& 732 & 51.20 \\
& & Single & 39.38 & 522 & - & 188 & 6.37 \\
& & \cellcolor{green!12}Improve
& \cellcolor{green!12}-0.25
& \cellcolor{green!12}48.2\%
& \cellcolor{green!12}99.6\%
& \cellcolor{green!12}85.5\%
& \cellcolor{green!12}7.1$\times$ \\
\cmidrule(rl){2-8}

& \multirow{5}{*}{Qwen3-14B}
& Ours & \textbf{53.67} & \underline{1466} & 3 & \textbf{169} & \textbf{13.30} \\
& & Latent-MAS & 51.13 & \textbf{1420} & 12 & \underline{196} & \underline{21.91} \\
& & Text-MAS & \underline{52.86} & 3141 & 888& 904 & 48.94 \\
& & Single & 49.67 & 406 & - & 71 & 3.73 \\
& & \cellcolor{green!12}Improve
& \cellcolor{green!12}+0.81
& \cellcolor{green!12}53.3\%
& \cellcolor{green!12}99.7\%
& \cellcolor{green!12}81.3\%
& \cellcolor{green!12}3.7$\times$ \\

\cmidrule(rl){2-8}

& \multirow{5}{*}{Mistral-7B}
& Ours & \underline{38.84} & 2021.45 & 3 & 301 & 16 \\
& & Latent-MAS & \textbf{38.95} & \underline{2037} & 12 & \underline{587} & \underline{38} \\
& & Text-MAS & 38.56 & \textbf{4647} & 1041& \textbf{1165} & \textbf{71} \\
& & Single & 36.61 & 597 & - & 224 & 7 \\
& & \cellcolor{green!12}Improve
& \cellcolor{green!12}+0.28
& \cellcolor{green!12}56.50%
& \cellcolor{green!12}99.7\%
& \cellcolor{green!12}74.16%
& \cellcolor{green!12}4$\times$ \\
\bottomrule
\end{tabular}%
}
\end{table*}


\begin{table*}[t]
\centering
\small
\setlength{\tabcolsep}{4pt}
\renewcommand{\arraystretch}{0.92}
\caption{Main performance comparison on all models across benchmarks (continued; Part 3/6: GSM8K, MedQA(ML)).}
\label{tab:full_main_results_part3}
\resizebox{\textwidth}{!}{%
\begin{tabular}{@{}p{1.5cm}p{1.9cm}lccccc@{}}
\toprule
\textbf{Dataset} & \textbf{Model} & \textbf{Method}
& \textbf{Acc.}
& \textbf{Text Tokens}
& \textbf{Comm. Passes}
& \textbf{AR Steps}
& \textbf{Latency}\\
\midrule
\multirow{20}{*}{\centering GSM8K}
& \multirow{5}{*}{Qwen2.5-7B}
& Ours & \underline{44.05} & \textbf{740} & 3 & \textbf{14} & \textbf{0.91} \\
& & Latent-MAS & \textbf{44.87} & \underline{787} & 12 & \underline{73} & \underline{6.34} \\
& & Text-MAS & 43.06 & 1823 & 530& 552 & 18.60 \\
& & Single & 31.84 & 372 & - & 195 & 8.26 \\
& & \cellcolor{green!12}Improve
& \cellcolor{green!12}+0.99
& \cellcolor{green!12}59.4\%
& \cellcolor{green!12}99.4\%
& \cellcolor{green!12}97.5\%
& \cellcolor{green!12}20.4$\times$ \\
\cmidrule(rl){2-8}

& \multirow{5}{*}{Qwen3-8B}
& Ours & 82.05 & \underline{755} & 3 & \underline{27} & \textbf{2.41} \\
& & Latent-MAS & \textbf{83.55} & \textbf{753} & 12 & \textbf{23} & \underline{4.28} \\
& & Text-MAS & \underline{83.45} & 1439 & 291& 296 & 20.07 \\
& & Single & 77.10 & 303 & - & 122 & 4.24 \\
& & \cellcolor{green!12}Improve
& \cellcolor{green!12}-1.40
& \cellcolor{green!12}47.5\%
& \cellcolor{green!12}99.0\%
& \cellcolor{green!12}90.9\%
& \cellcolor{green!12}8.3$\times$ \\
\cmidrule(rl){2-8}

& \multirow{5}{*}{Qwen3-14B}
& Ours & \underline{79.48} & \underline{933} & 3 & \textbf{148} & \textbf{9.99} \\
& & Latent-MAS & \textbf{82.13} & \textbf{785} & 12 & \underline{224} & 24.35 \\
& & Text-MAS & 78.54 & 1663 & 444& 450 & \underline{23.68} \\
& & Single & 71.20 & 296 & - & 111 & 5.67 \\
& & \cellcolor{green!12}Improve
& \cellcolor{green!12}+0.94
& \cellcolor{green!12}43.9\%
& \cellcolor{green!12}99.3\%
& \cellcolor{green!12}67.1\%
& \cellcolor{green!12}2.4$\times$ \\

\cmidrule(rl){2-8}

& \multirow{5}{*}{Mistral-7B}
& Ours & \underline{36.47} & \underline{1401.95} & 3 & 276 & 15 \\
& & Latent-MAS & \textbf{38.14} & 1390 & 12 & \underline{583} & \textbf{37} \\
& & Text-MAS & 36.45 & \textbf{2642} & 578& \textbf{600} & \underline{37} \\
& & Single & 25.09 & 329 & - & 124 & 5 \\
& & \cellcolor{green!12}Improve
& \cellcolor{green!12}+0.02
& \cellcolor{green!12}46.94%
& \cellcolor{green!12}99.5\%
& \cellcolor{green!12}54.00%
& \cellcolor{green!12}3$\times$ \\
\midrule
\multirow{20}{*}{\centering MedQA(ML)}
& \multirow{5}{*}{Qwen2.5-7B}
& Ours & \underline{82.96} & \textbf{882} & 3 & \textbf{5} & \textbf{0.74} \\
& & Latent-MAS & \textbf{85.65} & \underline{888} & 12 & \underline{20} & \underline{3.54} \\
& & Text-MAS & 81.95 & 1996 & 558& 569 & 15.86 \\
& & Single & 80.30 & 224 & - & 8 & 0.23 \\
& & \cellcolor{green!12}Improve
& \cellcolor{green!12}+1.01
& \cellcolor{green!12}55.8\%
& \cellcolor{green!12}99.5\%
& \cellcolor{green!12}99.5\%
& \cellcolor{green!12}21.4$\times$ \\
\cmidrule(rl){2-8}

& \multirow{5}{*}{Qwen3-8B}
& Ours & 80.12 & \underline{901} & 3 & \textbf{6} & \textbf{0.69} \\
& & Latent-MAS & \underline{80.18} & \textbf{897} & 12 & \underline{14} & \underline{3.04} \\
& & Text-MAS & \textbf{81.41} & 1930 & 514& 522 & 26.03 \\
& & Single & 77.05 & 294 & - & 74 & 3.70 \\
& & \cellcolor{green!12}Improve
& \cellcolor{green!12}-1.29
& \cellcolor{green!12}53.3\%
& \cellcolor{green!12}99.4\%
& \cellcolor{green!12}99.2\%
& \cellcolor{green!12}37.7$\times$ \\
\cmidrule(rl){2-8}

& \multirow{5}{*}{Qwen3-14B}
& Ours & 83.03 & \underline{952} & 3 & \textbf{96} & \textbf{7.68} \\
& & Latent-MAS & \underline{83.11} & \textbf{888} & 12 & \underline{104} & \underline{17.90} \\
& & Text-MAS & \textbf{84.68} & 2321 & 685& 693 & 36.88 \\
& & Single & 78.17 & 233 & - & 10 & 0.54 \\
& & Improve & -1.65 & 59.0\% & 99.6\% & 86.1\% & 4.8$\times$ \\

\cmidrule(rl){2-8}

& \multirow{5}{*}{Mistral-7B}
& Ours & \underline{46.93} & 1673.19 & 3 & 257 & 15 \\
& & Latent-MAS & \textbf{47.85} & \underline{1682} & 12 & \underline{585} & \underline{38} \\
& & Text-MAS & 41.41 & \textbf{3893} & 918& \textbf{975} & \textbf{61} \\
& & Single & 32.05 & 366 & - & 81 & 4 \\
& & \cellcolor{green!12}Improve
& \cellcolor{green!12}+5.52
& \cellcolor{green!12}57.02%
& \cellcolor{green!12}99.7\%
& \cellcolor{green!12}73.64%
& \cellcolor{green!12}4$\times$ \\
\bottomrule
\end{tabular}%
}
\end{table*}


\begin{table*}[htbp]
\centering
\small
\setlength{\tabcolsep}{4pt}
\renewcommand{\arraystretch}{0.92}
\caption{Main performance comparison on all models across benchmarks (continued; Part 4/6: MedQA(TW), MedQA(US)).}
\label{tab:full_main_results_part4}
\resizebox{\textwidth}{!}{%
\begin{tabular}{@{}p{1.5cm}p{1.9cm}lccccc@{}}
\toprule
\textbf{Dataset} & \textbf{Model} & \textbf{Method}
& \textbf{Acc.}
& \textbf{Text Tokens}
& \textbf{Comm. Passes}
& \textbf{AR Steps}
& \textbf{Latency}\\
\midrule
\multirow{20}{*}{\centering MedQA(TW)}
& \multirow{5}{*}{Qwen2.5-7B}
& Ours & \underline{60.20} & \textbf{960} & 3 & \textbf{5} & \textbf{0.86} \\
& & Latent-MAS & 59.46 & \underline{971} & 12 & \underline{26} & \underline{3.91} \\
& & Text-MAS & \textbf{62.88} & 2092 & 565& 580 & 19.85 \\
& & Single & 57.64 & 247 & - & 13 & 0.35 \\
& & Improve & -2.68 & 54.1\% & 99.5\% & 99.5\% & 23.1$\times$ \\
\cmidrule(rl){2-8}

& \multirow{5}{*}{Qwen3-8B}
& Ours & \textbf{73.41} & \underline{989} & 3 & \textbf{8} & \textbf{2.34} \\
& & Latent-MAS & 72.26 & \textbf{976} & 12 & \underline{15} & \underline{4.52} \\
& & Text-MAS & \underline{73.25} & 2162 & 589& 603 & 29.89 \\
& & Single & 65.96 & 336 & - & 97 & 6.01 \\
& & \cellcolor{green!12}Improve
& \cellcolor{green!12}+0.16
& \cellcolor{green!12}54.3\%
& \cellcolor{green!12}99.5\%
& \cellcolor{green!12}98.7\%
& \cellcolor{green!12}12.8$\times$ \\
\cmidrule(rl){2-8}

& \multirow{5}{*}{Qwen3-14B}
& Ours & \underline{82.76} & \underline{1032} & 3 & \textbf{102} & \textbf{8.54} \\
& & Latent-MAS & \textbf{82.93} & \textbf{976} & 12 & \underline{118} & \underline{18.33} \\
& & Text-MAS & 81.74 & 2375 & 669& 678 & 36.13 \\
& & Single & 77.64 & 256 & - & 15 & 0.80 \\
& & \cellcolor{green!12}Improve
& \cellcolor{green!12}+1.02
& \cellcolor{green!12}56.5\%
& \cellcolor{green!12}99.6\%
& \cellcolor{green!12}85.0\%
& \cellcolor{green!12}4.2$\times$ \\

\cmidrule(rl){2-8}

& \multirow{5}{*}{Mistral-7B}
& Ours & \underline{46.84} & 1778.75 & 3 & 270 & 18 \\
& & Latent-MAS & \textbf{47.35} & \underline{1802} & 12 & \underline{588} & \underline{39} \\
& & Text-MAS & 45.78 & \textbf{4146} & 982& \textbf{1018} & \textbf{63} \\
& & Single & 41.61 & 410 & - & 92 & 5 \\
& & \cellcolor{green!12}Improve
& \cellcolor{green!12}+1.06
& \cellcolor{green!12}57.10%
& \cellcolor{green!12}99.7\%
& \cellcolor{green!12}73.48%
& \cellcolor{green!12}4$\times$ \\
\midrule
\multirow{20}{*}{\centering MedQA(US)}
& \multirow{5}{*}{Qwen2.5-7B}
& Ours & 45.25 & \textbf{1432} & 3 & \textbf{7} & \textbf{1.29} \\
& & Latent-MAS & \textbf{46.20} & \underline{1438} & 12 & \underline{25} & \underline{4.21} \\
& & Text-MAS & \underline{45.48} & 2752 & 665& 676 & 23.09 \\
& & Single & 44.27 & 361 & - & 10 & 0.29 \\
& & \cellcolor{green!12}Improve
& \cellcolor{green!12}-0.23
& \cellcolor{green!12}48.0\%
& \cellcolor{green!12}99.5\%
& \cellcolor{green!12}99.0\%
& \cellcolor{green!12}17.9$\times$ \\
\cmidrule(rl){2-8}

& \multirow{5}{*}{Qwen3-8B}
& Ours & 71.16 & \textbf{1421} & 3 & \textbf{14} & \textbf{3.11} \\
& & Latent-MAS & \textbf{77.34} & \underline{1438} & 12 & \underline{16} & \underline{3.51} \\
& & Text-MAS & \underline{72.53} & 2552 & 548& 561 & 24.70 \\
& & Single & 55.12 & 521 & - & 166 & 10.43 \\
& & \cellcolor{green!12}Improve
& \cellcolor{green!12}-1.37
& \cellcolor{green!12}44.3\%
& \cellcolor{green!12}99.5\%
& \cellcolor{green!12}97.5\%
& \cellcolor{green!12}7.9$\times$ \\
\cmidrule(rl){2-8}

& \multirow{5}{*}{Qwen3-14B}
& Ours & \underline{69.34} & \underline{1494} & 3 & \textbf{99} & \textbf{9.44} \\
& & Latent-MAS & 68.18 & \textbf{1451} & 12 & \underline{128} & \underline{18.44} \\
& & Text-MAS & \textbf{69.99} & 3091 & 803& 812 & 43.98 \\
& & Single & 62.22 & 368 & - & 10 & 0.60 \\
& & \cellcolor{green!12}Improve & \cellcolor{green!12}-0.65 & \cellcolor{green!12}51.7\% & \cellcolor{green!12}99.6\% & \cellcolor{green!12}87.8\% & \cellcolor{green!12}4.7$\times$ \\
\cmidrule(rl){2-8}

& \multirow{5}{*}{Mistral-7B}
& Ours & \underline{42.53} & 2135.86 & 3 & 284 & 17 \\
& & Latent-MAS & \textbf{43.53} & \underline{2200} & 12 & \underline{586} & \underline{39} \\
& & Text-MAS & 41.15 & \textbf{4957} & 1045& \textbf{1134} & \textbf{61} \\
& & Single & 39.20 & 527 & - & 119 & 5 \\
& & \cellcolor{green!12}Improve
& \cellcolor{green!12}+1.38
& \cellcolor{green!12}56.91%
& \cellcolor{green!12}99.7\%
& \cellcolor{green!12}74.96%
& \cellcolor{green!12}4$\times$ \\
\bottomrule
\end{tabular}%
}
\end{table*}


\begin{table*}[htbp]
\centering
\small
\setlength{\tabcolsep}{4pt}
\renewcommand{\arraystretch}{0.92}
\caption{Main performance comparison on all models across benchmarks (continued; Part 5/6: OpenBooKQA, MMLU).}
\label{tab:full_main_results_part5}
\resizebox{\textwidth}{!}{%
\begin{tabular}{@{}p{1.5cm}p{1.9cm}lccccc@{}}
\toprule
\textbf{Dataset} & \textbf{Model} & \textbf{Method}
& \textbf{Acc.}
& \textbf{Text Tokens}
& \textbf{Comm. Passes}
& \textbf{AR Steps}
& \textbf{Latency}\\
\midrule
\multirow{20}{*}{\centering OpenBooKQA}
& \multirow{5}{*}{Qwen2.5-7B}
& Ours & \underline{77.60} & \textbf{679} & 3 & \textbf{4} & \textbf{0.84} \\
& & Latent-MAS & 75.80 & \underline{681} & 12 & \underline{16} & \underline{3.24} \\
& & Text-MAS & \textbf{80.40} & 1646 & 479& 486 & 17.02 \\
& & Single & 71.42 & 176 & - & 7 & 0.23 \\
& & Improve
& -2.80
& 58.7\%
& 99.4\%
& 99.6\%
& 20.3$\times$ \\
\cmidrule(rl){2-8}

& \multirow{5}{*}{Qwen3-8B}
& Ours & \underline{86.23} & \textbf{687} & 3 & \textbf{6} & \textbf{1.21} \\
& & Latent-MAS & \textbf{87.61} & \underline{696} & 12 & \underline{15} & \underline{4.18} \\
& & Text-MAS & 86.12 & 1510 & 395& 403 & 14.36 \\
& & Single & 85.45 & 219 & - & 50 & 1.78 \\
& & \cellcolor{green!12}Improve
& \cellcolor{green!12}+0.11
& \cellcolor{green!12}54.5\%
& \cellcolor{green!12}99.2\%
& \cellcolor{green!12}99.0\%
& \cellcolor{green!12}11.9$\times$ \\
\cmidrule(rl){2-8}

& \multirow{5}{*}{Qwen3-14B}
& Ours & \underline{92.73} & \underline{767} & 3 & \textbf{99} & \textbf{7.86} \\
& & Latent-MAS & \textbf{95.85} & \textbf{705} & 12 & \underline{109} & \underline{8.41} \\
& & Text-MAS & 91.62 & 1592 & 438& 446 & 23.41 \\
& & Single & 87.14 & 183 & - & 10 & 0.50 \\
& & \cellcolor{green!12}Improve
& \cellcolor{green!12}+1.11
& \cellcolor{green!12}51.8\%
& \cellcolor{green!12}99.3\%
& \cellcolor{green!12}77.8\%
& \cellcolor{green!12}3.0$\times$ \\
\cmidrule(rl){2-8}

& \multirow{5}{*}{Mistral-7B}
& Ours & \underline{72.41} & 1325.71 & 3 & 289 & 11 \\
& & Latent-MAS & \textbf{73.54} & \underline{1338} & 12 & \underline{579} & \underline{25} \\
& & Text-MAS & 70.65 & \textbf{3066} & 784& \textbf{846} & \textbf{51} \\
& & Single & 62.82 & 237 & - & 46 & 1 \\
& & \cellcolor{green!12}Improve
& \cellcolor{green!12}+1.76
& \cellcolor{green!12}56.77%
& \cellcolor{green!12}99.6\%
& \cellcolor{green!12}65.84%
& \cellcolor{green!12}5$\times$ \\
\midrule
\multirow{20}{*}{\centering MMLU}
& \multirow{5}{*}{Qwen2.5-7B}
& Ours & 64.38 & \textbf{946} & 3 & \textbf{5} & \textbf{0.88} \\
& & Latent-MAS & \textbf{65.93} & \underline{952} & 12 & \underline{21} & \underline{2.65} \\
& & Text-MAS & \underline{64.88} & 1934 & 513& 526 & 17.68 \\
& & Single & 62.01 & 244 & - & 12 & 0.69 \\
& & \cellcolor{green!12}Improve
& \cellcolor{green!12}-0.50
& \cellcolor{green!12}51.1\%
& \cellcolor{green!12}99.4\%
& \cellcolor{green!12}99.4\%
& \cellcolor{green!12}20.1$\times$ \\
\cmidrule(rl){2-8}

& \multirow{5}{*}{Qwen3-8B}
& Ours & \textbf{82.79} & \textbf{956} & 3 & \underline{22} & \textbf{2.66} \\
& & Latent-MAS & 81.47 & \underline{1010} & 12 & \textbf{16} & \underline{2.97} \\
& & Text-MAS & \underline{82.55} & 1887 & 439& 455 & 24.75 \\
& & Single & 75.34 & 316 & - & 108 & 4.36 \\
& & \cellcolor{green!12}Improve
& \cellcolor{green!12}+0.24
& \cellcolor{green!12}49.3\%
& \cellcolor{green!12}99.3\%
& \cellcolor{green!12}95.2\%
& \cellcolor{green!12}9.3$\times$ \\
\cmidrule(rl){2-8}

& \multirow{5}{*}{Qwen3-14B}
& Ours & 83.18 & \textbf{1001} & 3 & \textbf{27} & \textbf{3.04} \\
& & Latent-MAS & \textbf{84.19} & \underline{1056} & 12 & \underline{29} & \underline{3.47} \\
& & Text-MAS & \underline{83.93} & 1899 & 415& 471 & 25.75 \\
& & Single & 79.58 & 329 & - & 114 & 4.67 \\
& & \cellcolor{green!12}Improve
& \cellcolor{green!12}-0.75
& \cellcolor{green!12}47.3\%
& \cellcolor{green!12}99.3\%
& \cellcolor{green!12}94.3\%
& \cellcolor{green!12}8.5$\times$ \\
\cmidrule(rl){2-8}

& \multirow{5}{*}{Mistral-7B}
& Ours & 56.28 & 1422.48 & 3 & 263 & 17 \\
& & Latent-MAS & \underline{56.56} & \underline{1439} & 12 & \underline{583} & \underline{39} \\
& & Text-MAS & \textbf{57.53} & \textbf{3862} & 981& \textbf{1052} & \textbf{58} \\
& & Single & 48.95 & 337 & - & 70 & 4 \\
& & \cellcolor{green!12}Improve
& \cellcolor{green!12}-1.25
& \cellcolor{green!12}63.17%
& \cellcolor{green!12}99.7\%
& \cellcolor{green!12}75.00%
& \cellcolor{green!12}3$\times$ \\
\bottomrule
\end{tabular}%
}
\end{table*}


\begin{table*}[htbp]
\centering
\small
\setlength{\tabcolsep}{4pt}
\renewcommand{\arraystretch}{0.92}
\caption{Main performance comparison on all models across benchmarks (continued; Part 6/6: QuAIL).}
\label{tab:full_main_results_part6}
\resizebox{\textwidth}{!}{%
\begin{tabular}{@{}p{1.5cm}p{1.9cm}lccccc@{}}
\toprule
\textbf{Dataset} & \textbf{Model} & \textbf{Method}
& \textbf{Acc.}
& \textbf{Text Tokens}
& \textbf{Comm. Passes}
& \textbf{AR Steps}
& \textbf{Latency}\\
\midrule
\multirow{20}{*}{\centering QuAIL}
& \multirow{5}{*}{Qwen2.5-7B}
& Ours & 67.07 & \underline{2388} & 3 & \textbf{9} & \textbf{1.74} \\
& & Latent-MAS & \textbf{71.87} & \textbf{2387} & 12 & \underline{19} & \underline{4.04} \\
& & Text-MAS & \underline{68.88} & 3066 & 312& 336 & 11.97 \\
& & Single & 65.60 & 599 & - & 7 & 0.32 \\
& & Improve
& -1.81
& 22.1\%
& 99.0\%
& 97.3\%
& 6.9$\times$ \\
\cmidrule(rl){2-8}

& \multirow{5}{*}{Qwen3-8B}
& Ours & \underline{77.44} & \textbf{2378} & 3 & \textbf{6} & \textbf{1.49} \\
& & Latent-MAS & 77.36 & \underline{2498} & 12 & \underline{14} & \underline{5.33} \\
& & Text-MAS & \textbf{78.66} & 3107 & 311& 350 & 12.36 \\
& & Single & 72.98 & 612 & - & 17 & 0.68 \\
& & \cellcolor{green!12}Improve
& \cellcolor{green!12}-1.22
& \cellcolor{green!12}23.5\%
& \cellcolor{green!12}99.0\%
& \cellcolor{green!12}98.9\%
& \cellcolor{green!12}8.3$\times$ \\
\cmidrule(rl){2-8}

& \multirow{5}{*}{Qwen3-14B}
& Ours & \textbf{80.57} & \underline{2474} & 3 & \textbf{101} & \textbf{12.70} \\
& & Latent-MAS & 79.47 & \textbf{2468} & 12 & \underline{167} & \underline{19.06} \\
& & Text-MAS & \underline{79.91} & 3278 & 431& 439 & 24.38 \\
& & Single & 77.26 & 608 & - & 8 & 0.59 \\
& & \cellcolor{green!12}Improve
& \cellcolor{green!12}+0.66
& \cellcolor{green!12}24.5\%
& \cellcolor{green!12}99.3\%
& \cellcolor{green!12}77.0\%
& \cellcolor{green!12}1.9$\times$ \\
\cmidrule(rl){2-8}

& \multirow{5}{*}{Mistral-7B}
& Ours & \underline{61.43} & 3121.59 & 3 & 296 & 14 \\
& & Latent-MAS & \textbf{62.94} & \underline{3139} & 12 & \underline{580} & \underline{36} \\
& & Text-MAS & 61.15 & \textbf{4317} & 584& \textbf{620} & \textbf{37} \\
& & Single & 59.25 & 673 & - & 24 & 1 \\
& & \cellcolor{green!12}Improve
& \cellcolor{green!12}+0.28
& \cellcolor{green!12}27.69%
& \cellcolor{green!12}99.5\%
& \cellcolor{green!12}52.26%
& \cellcolor{green!12}3$\times$ \\
\bottomrule
\end{tabular}%
}
\end{table*}


\clearpage

\begin{figure}[htbp]
  \centering
  \includegraphics[width=1\textwidth]{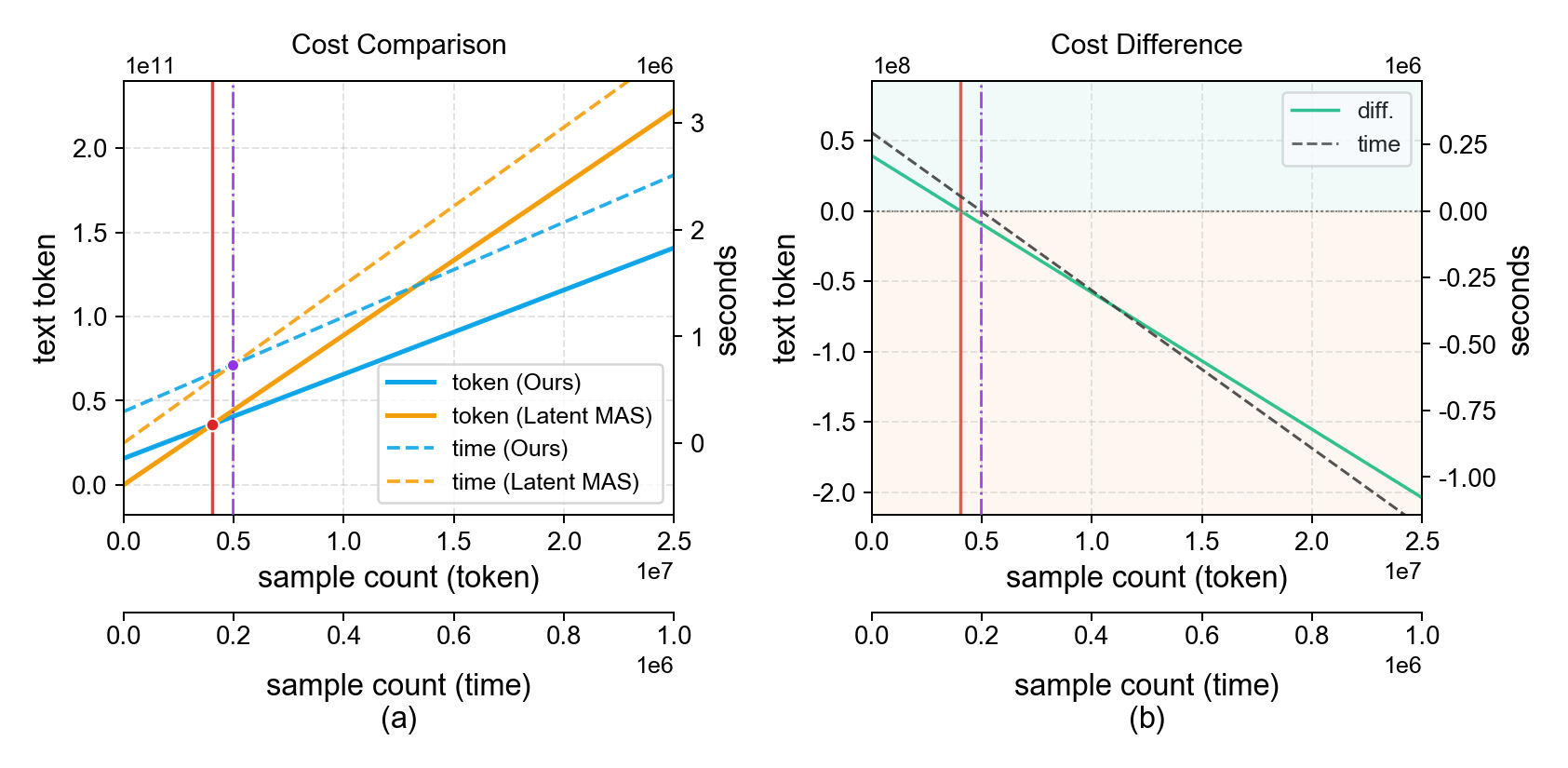}
  \caption{Cumulative token and latency scaling of our method versus autoregressive latent MAS on Qwen3-8B.
  The left panels show cumulative token consumption and latency, and the right panels show the corresponding differences, computed as ours minus latent MAS.}
  \label{fig:token_and_time}
\end{figure}

\paragraph{Amortized Efficiency at Scale.}
\label{app:amortized_efficiency}
Fig.~\ref{fig:token_and_time} compares the cumulative token and latency cost of Thought Blocks against autoregressive latent MAS on Qwen3-8B.
The x-axis denotes the number of task instances under our default four-agent setting.
Although Thought Blocks introduce an additional one-time training cost, this cost is fixed with respect to deployment volume, whereas the inference savings accumulate over every subsequent multi-agent call.
Under this setting, our method becomes more token-efficient after approximately \(8.82\times10^4\) task instances and faster after approximately \(3.37\times10^5\) task instances.

These crossover points should be interpreted as conservative for more communication-intensive workflows.
Our experiments use four agents per task.
If a task requires more agents, more communication rounds, or richer inter-agent state transfer, the per-instance saving from removing autoregressive latent rollout increases, while the one-time training cost remains unchanged.
Therefore, the amortization point would move earlier as the amount of inter-agent communication per task grows.

The task counts required for amortization may appear large in a standalone benchmark setting, but they are modest in shared deployment scenarios.
General-purpose AI systems already serve hundreds of millions of weekly users, and recent enterprise surveys report rapidly growing experimentation with AI agents~\citep{bellan2025chatgpt800m,mckinsey2025stateai}.
We do not assume that current deployments exactly match our experimental setting.
Rather, these trends suggest that the cost of inter-agent communication will become increasingly important as agentic systems move from small closed multi-agent teams toward higher-volume services and cross-agent interactions, where agents may need to represent different users, tools, or organizations.
In such settings, reducing both sequential inference cost and communication memory traffic can translate into substantial aggregate savings.

\clearpage
\subsection{Implementation Details}
\label{app:implementation_details}

\paragraph{Hardware Environment.}
All experiments are conducted on a server running CentOS Linux 7.
The server is equipped with two Intel Xeon Gold 6430 CPUs and eight NVIDIA RTX A6000 GPUs with 48 GB memory each.
Unless otherwise specified, Writer training and latency evaluation use a single visible GPU rather than distributed data-parallel execution.

\paragraph{Beacon Backbone Training.}
We build the compressed-context backbone following Activation Beacon~\citep{zhang2024activation} and follow the original configuration without modification.
During Beacon training, the language-model backbone is frozen and only the Beacon parameters are optimized, including the Beacon embedding and the Beacon-specific query, key, and value projections.
Beacon positions are interleaved with the original tokens; both the sliding window and stride are set to 2,048 tokens, with full-coverage attention and access to previous Beacon states.

Beacon training consists of pretraining followed by long-context fine-tuning.
For pretraining, we follow the Activation Beacon recipe based on the RedPajama-Data-1T-Sample corpus.
Each document is terminated with an EOS token and documents are not packed together.
Only the Beacon-specific parameters are optimized during this stage.

For long-context fine-tuning, we use a data mixture, consisting of GPT-generated long-document QA from books and papers, LongAlpaca, BookSum, synthetic needle-retrieval data, and a 5K-example RedPajama subset.
We use a maximum sequence length of 8,192 tokens, a minimum sequence length of 1,024 tokens, one fine-tuning epoch, a learning rate of $1\times10^{-5}$, bf16 precision, FlashAttention-2, and DeepSpeed ZeRO-2.

\paragraph{Teacher State Construction.}
The pre-trajectory continuation state is constructed using the same
state representation as in the main method: earlier groups retain only
Beacon KV states, while the final group keeps its raw-token KV states,
and the layer-wise hidden states at the final prefill position form the
continuation seed.

For trajectory-to-state supervision, the frozen teacher performs
autoregressive latent decoding from the pre-trajectory state
$(R_i,C_i)$. The teacher rollout follows a fixed decoding horizon,
capped at 12 decode steps under the standard configuration, with an
extended 96-step configuration used only for larger slot-count studies.
Decode-stage Beacons are placed along the resulting trajectory.

After constructing the teacher rollout, the state-extraction operator
$\Pi$ converts the fixed-horizon latent trajectory into a fixed-size
teacher continuation-memory target $\mathcal{M}_i^\star$, represented
by $B$ Beacon states. Here, $B$ denotes the size of the distilled
continuation-memory representation rather than a truncation of the
teacher rollout.
 
The retained states are model-native continuation states reached along
the teacher's autoregressive latent trajectory. The Thought Block writer
is trained to directly reconstruct the corresponding continuation-memory
target from $(R_i,C_i)$, replacing the sequential autoregressive
construction of these states with a fixed-size one-shot prediction.
The teacher trajectory is therefore required only for constructing
supervision during training and is removed at inference time.

\paragraph{Thought Block Writer Training.}
The language-model backbone and Beacon mechanism remain frozen throughout this stage, and only the Writer parameters $\phi$ are optimized.
Our standard configuration uses $B=4$ learnable slots and a Beacon compression ratio of $r=2$. For each teacher rollout, we collect the valid decode-stage Beacon states in their causal order and retain the $B$ such states as the fixed-size continuation-memory target $\mathcal{M}_i^\star$. Importantly, the block size $B$ determines the number of
retained Beacon states rather than the length
of the teacher rollout. The teacher decoding
horizon is configured to provide sufficient
valid Beacon positions for the corresponding
block size. Writer training is performed purely through latent-state distillation.

We train the Writer with AdamW for 10 epochs using a learning rate of
$5\times10^{-5}$, weight decay of $10^{-2}$, gradient clipping at 1.0,
and dropout 0.
The training loss combines KV-state matching and the auxiliary
hidden-state alignment defined in Eq.~\ref{eq:app_aux_loss}, with
weights $\lambda_{\mathrm{KV}}=1.0$ and
$\lambda_{\mathrm{aux}}=0.5$, respectively.

The mixed teacher-trace dataset
contains 78,207 valid training examples.
We use a batch size of 8, corresponding to 9,776 optimization steps per epoch. The training examples consist of synthetic data and the training splits of the benchmarks, including ARC-Easy, BoolQ, GSM8K, GPQA,
MedQA-US, MedQA-Mainland, MedQA-Taiwan, MMLU, OpenBookQA, and QuAIL. Each training example records the role prompt, the pre-trajectory continuation representation, and the hidden/KV states associated with decode-stage Beacon positions. All reported benchmark results are evaluated on the evaluation/test splits, and no
test-split teacher traces are used for Writer training.

\paragraph{Latency and Efficiency Measurement.}
End-to-end latency is measured sample-by-sample with an effective
inference batch size of 1.
The timer covers the complete multi-agent inference pipeline, from the
first agent's initial prefill to completion of the final output,
including prompt construction, tokenization, model forward passes,
inter-agent state-transfer logic, and final decoding. We measure host-observed wall-clock time using
\texttt{time.perf\_counter} and report the total elapsed time divided by the number of evaluated samples as the average per-sample latency. We call \texttt{torch.cuda.synchronize()} immediately before starting and after completing each timed inference to ensure that all queued CUDA kernels are included in the wall-clock measurement.

The reported KV-cache load is computed analytically from the
corresponding KV tensors using their shapes, element sizes, and data
types.
It is therefore a memory-volume proxy rather than a hardware-profiler
measurement of actual PCIe or HBM traffic.




\end{document}